\documentclass[article]{aa}
\usepackage{natbib}
\usepackage[utf8]{inputenc}
\usepackage{txfonts}
\usepackage{graphicx}
\usepackage{datetime}
\usepackage{color}

\usepackage{hyperref}
\hypersetup{
        colorlinks = true,
        urlcolor = blue,
        citecolor=blue
}

\begin{document}

\title{\bf The magnetic connectivity of coronal shocks from behind-the-limb flares to the visible solar surface during $\gamma$-ray events}
%\subtitle{}

\titlerunning{Solar $\gamma$-rays during far-side CMEs}

\author{I. Plotnikov \inst{1,2}, A .P.  Rouillard\inst{1,2}, \and G.H. Share\inst{3,4}}
  
\institute{Universite de Toulouse; UPS-OMP; IRAP;  Toulouse, France
   \and
  CNRS; IRAP; 9 Av. colonel Roche, BP 44346, F-31028 Toulouse cedex 4, France
    \and
    Department of Astronomy, University of Maryland, College Park, MD 20742, USA
    \and
    National Observatory of Athens, P.O. Box 20048, Thissio, 11810 - Athens, Greece\\
Contact: \email{illya.plotnikov@irap.omp.eu}}

%\date{\emph{manuscript version:} \today; \currenttim}
\date{}

% \abstract{}{}{}{}{} 
% 5 {} token are mandatory
 
\abstract
%context (optional)
{The observation of $>$100 MeV $\gamma$-rays in the minutes to hours following solar flares suggests that high-energy particles interacting in the solar atmosphere can be stored and/or accelerated for long time periods.  The occasions when $\gamma$-rays are detected even when the solar eruptions occurred beyond the solar limb as viewed from Earth provide favorable viewing conditions for studying the role of coronal shocks driven by coronal mass ejections (CMEs) in the acceleration of these particles. }
%
% aims 
{In this paper, we investigate the spatial and temporal evolution of the coronal shocks inferred from stereoscopic observations of  behind-the-limb flares to determine if they could be the source of the particles producing the $\gamma$-rays.}
%
%methods 
{We analyzed the CMEs and early formation of coronal shocks associated with $\gamma-$ray events measured by the {\it Fermi}-Large Area Telescope (LAT) from three eruptions behind the solar limb as viewed from Earth on 2013 Oct 11, 2014 Jan 6 and Sept 1.  We used a 3D triangulation technique, based on remote-sensing observations to model the expansion of the CME shocks from above the solar surface to the upper corona.  Coupling the expansion model to various models of the coronal magnetic field allowed us to derive the time-dependent distribution of shock Mach numbers and the magnetic connection of particles produced by the shock to the solar surface visible from Earth. }
%
%results
{The reconstructed shock fronts for the three events became magnetically connected to the visible solar surface after the start of the flare and just before the onset of the $>$ 100 MeV $\gamma$-ray emission. The shock surface at these connections also exhibited supercritical Mach numbers required for significant particle energization. The strongest $\gamma$-ray emissions occurred when the flanks of the shocks were connected in a quasi-perpendicular geometry to the field lines reaching the visible surface. Multipoint, in situ, measurements of solar energetic particles (SEPs) were consistent with the production of these SEPs by the same shock processes responsible for the $\gamma$-rays.   The fluxes of protons in space and at the Sun were highest for the 2014 Sept 1, which had the fastest moving shock.}
%
% conclusion 
{This study provides further evidence that high-energy protons producing time-extended high-energy $\gamma$-ray emission  likely have the same CME-shock origin as solar energetic particles measured in interplanetary space.}
  
  \keywords{  Sun: flares -- Sun: X-rays, gamma rays  -- Sun: coronal mass ejections (CMEs) - Sun: magnetic fields }
  
  \maketitle
  
  \section{Introduction}
Observations of $\gamma$-rays and neutrons during flares provide the only information on the properties of very energetic ions accelerated in the corona and impacting the solar atmosphere \citep{2011SSRv..159..167V}. Energetic ions interacting with the solar atmosphere produce a wealth of $\gamma$-ray emissions. A $\gamma$-ray line spectrum is produced through interactions of ions in the range 1 –- 100 MeV/nucleon and consists of several nuclear de-excitation lines, neutron capture, and positron annihilation lines. If more energetic ions are present, over a few hundred
MeV/nucleon, nuclear interactions with the ambient medium produce secondary pions whose decay products lead to a broadband continuum detectable at photon energies above 10 MeV. Such high-energy $\gamma$-ray emission in solar flares was widely studied since early 80s \citep[e.g.,][]{1981ICRC...10....5F, 1987ApJS...63..721M, 1987ApJ...316L..41R}. The first $\gamma$-ray line images of a solar flare were obtained with the Reuven Ramaty High Energy Solar Spectroscopic Imager \citep[\textit{RHESSI};][]{2002SoPh..210....3L} for the X4.8 flare of 2002 July 23 \citep{2003ApJ...595L..77H}.

 For some of the events, high-energy emission has been observed for hours after the impulsive phase of the flare, revealing that high-energy ions must be continuously accelerated or stored for tens of minutes to hours \citep[e.g.,][]{1993A&AS...97..349K, 1996SoPh..166..107A}. These high-energy events were named long duration gamma ray flares \citep[LDGRFs;][]{Ryan00}. If particle, responsible for this $\gamma$-ray emission were to be accelerated in the flare region \citep[flare scenario; e.g.,][]{2012SSRv..171....3A} then the emission should be confined to the proximity of the eruptive loop. Competing particle acceleration mechanisms in flares include acceleration in the reconnection current sheet \citep{1996ApJ...462..997L}, stochastic acceleration by turbulence \citep{2012SSRv..173..535P}, and flare termination shocks \citep{1985ApJ...298..400E, 2015Sci...350.1238C}.  The LDGRFs occurring for several hours challenge the flare hypothesis since emissions can outlive the solar flares and trapping of $>$100MeV protons on coronal loops seems theoretically unlikely. This class of time-extended $\gamma$-ray events often show pion-decay radiation over tens of minutes to hours after the impulsive phase, while other common flare emissions (e.g., X-rays) are absent or greatly diminished \citep[e.g.,][]{2008LRSP....5....1B, 2011SSRv..158....5H}. Share et al. (2017) (to be submitted) have completed a study 29 of these time-extended $\gamma$-ray events observed by LAT from the 2008 to 2016.  As the events are distinct from the accompanying solar flares and have varying durations, these authors have named them sustained $\gamma$-ray emission (SGRE) events.  We use this nomenclature in the remainder of this paper.

 The hadronic processes can result from the interaction of very energetic protons with the collisional chromospheric region situated between the photosphere and the weakly collisional corona where efficient particle acceleration occurs in either solar flares or coronal shocks. For this interpretation, $\gamma-$ray events that are detected near Earth require that energetic protons have propagated and impacted the chromospheric regions visible from the Earth (the so-called visible disk). Indeed, an additional constraint on the size of the accelerator and particle transport is imposed by the detection of SGRE from the events launched behind-the-limb. For instance an accelerator limited to the vicinity of the flare loop top is expected to influence a smaller region of the corona than a shock wave generated along a CME front.

 With the launch of the {\it Fermi} satellite many more long-duration $>$100 MeV events have been detected  \citep{2012ApJS..203....4A} that are  associated with much weaker X-ray flares. {\it Fermi}/Large Area Telescope (LAT) is a pair-conversion telescope \citep{atwo09} that is sensitive to $\gamma$-rays from about 20 MeV to 300 GeV. In the standard {\it{Fermi}} sky-survey mode the detector axis rocks $\pm$50$^{\circ}$ from the zenith each orbit; therefore, at most times during the year the Sun can be observed for $\sim$20-40 min every two orbits. The large aperture (2.4 sr) and effective area of the LAT provide the capability to sensitively monitor the Sun with a duty cycle of 15-20\%.  \citet{2012ApJS..203....4A} presented a detailed description of the solar flare analysis procedures and instrument response functions for the telemetered LAT data. A comprehensive analysis of LAT solar events is provided in a paper in preparation (Share et al. 2017, to be submitted). In the present study, we use the publicly available Pass7 and Pass8 source-class data.   The source-class data are used for most celestial source analyses for which long-time exposures are required.  Our method for analyzing the LAT data is best described as a 'light bucket' approach and is less sophisticated and faster than the maximum likelihood approach used by \citet{acke14, ajel14}, but it is almost as sensitive for detecting long-duration high-energy solar $\gamma$-ray events. With this approach we reduced the $\gamma$-ray background from the atmosphere of the Earth by restricting the allowable events to those with zenith angles $<$100$^{\circ}$.  We also restricted our studies to $\gamma$-rays with energies $>$100 MeV.  With these restrictions we accumulated all photons that have measured locations within about 10$^{\circ}$ of the Sun, thus basically a `light bucket'.  About 95\% of all 200 MeV solar $\gamma$-rays have measured locations within this 10$^{\circ}$ region of interest (ROI).

 \citet{2015ApJ...805L..15P, 2015arXiv150704303P} and \citet{2017ApJ...835..219A} have identified three $\gamma$-ray events observed by LAT that occured during behind-the-limb flare events. Because the 2014 Sept 01 occured well beyond the limb (36 degrees), the observed $\gamma$-ray emission could not have come from the flare footprints or loop top. Thus one needs to consider an acceleration process that can explain such an observation.  Acceleration at a coronal shock is a natural candidate for broadly distributed particle acceleration\footnote{Another possible scenario involves long-lasting post-CME coronal energy release \citep[e.g.,][]{1996SoPh..166..107A, 1996R&QE...39..940C, 2001SSRv...95..215K, 2014A&A...572A...4K}.}. It was proposed by \citet{Cliver93} that $\gamma-$ray events during solar events launched behind-the-limb are the consequence of particles acceleration by CME-driven coronal shocks in conjunction with their precipitation on the visible disk. The interaction of these particles with the dense solar plasma would then produce at least a part of the observed $\gamma-$ray flux.
 
 During its propagation out into the high corona and further into interplanetary space, the CME shock transfers a part of its kinetic energy to energetic particles; this typically amounts to roughly 10\% \citep{Mewaldt08}. These particles can eventually reach by number (flux greater than 10 particle cm$^{-2}$s$^{-1}$sr$^{-1}$) and energy ($\geq 10$~MeV) the levels required to be identified as a solar energetic particle event (SEP) at 1~AU \citep{Gopalswamy03}. It has been argued that most gradual SEP events observed in the interplanetary space are accelerated at the shock driven by the CME \citep{1999SSRv...90..413R, 2016ApJ...832..128C}. The highest particle energization (beyond GeV energies) occurs during the acceleration phase of the shock in the low corona \citep{2000JGR...10525079Z, 2003AstL...29..530B, 2005ApJS..158...38L, 2015A&A...584A..81A} when the shock can reach speeds up to $3000$~km/s and high Alfv\'enic Mach numbers $M_A \gg 1$, which are necessary for the efficient shock acceleration \citep[e.g.,][]{1987PhR...154....1B, 2016RPPh...79d6901M}. In the extreme example of astrophysical shocks where $M_A$ reaches very high values above $10^3$ (i.e., shocks of supernova remnants) there is evidence that protons are accelerated up to the knee of the cosmic ray spectrum, i.e., $10^{15}$~eV \citep{2012A&A...538A..81M}. In the coronal shocks the Mach numbers are more modest, typically 2-4, and probably do not exceed $\sim 20$ \citep{Rouillard16} and the available acceleration time is bound by the shock transit time of several hours to days. When the shock reaches the upper corona it decelerates and passes through more tenuous regions of the corona, diminishing the number and maximum energy of the accelerated particles. It was demonstrated by \citet{2008ApJ...686L.123N} that a 2500~km/s shock in a typical coronal conditions can accelerate protons up to $\sim$GeV energies in 10~minutes. Also, the shock acceleration scenario is supported by in situ energetic particle observations (albeit at large heliocentric distances) when the shock of a CME or a corotating interaction region impacts an observing spacecraft. 

Other researchers have used SDO observations to model the expanding coronal shock and its interaction with open field lines onto which particles would be released.  These studies allowed comparisons with SEPs observed in interplanetary space.  \citet{2015ApJ...799..167K} developed methods similar to those in the present study and combined a geometric model of the expanding shock front with the PFSS model of the coronal magnetic field.  \citet{2014ApJ...797....8L, 2017ApJ...838...51L} used a slightly different multipoint 3D reconstruction method than ours and used both PFSS and MHD MAST models to derive magnetic connectivity between the Sun and 1~AU to  understand the longitudinal spread of SEPs at 1~AU; both models are presented in detail in Sections~\ref{subsect:pfss} and \ref{subsect:mast} of this work.  Neither of these studies addressed the acceleration of particles that return to interact in the atmosphere.

 The SEPs measured near 1~AU could be the interplanetary counterpart, i.e., the anti-sunward rather than the sunward propagating particles, of the SGRE events. Therefore the flux and the maximal energy of SEPs at 1~AU (discussed in the section~\ref{sect:sep_obs}) could provide additional clues regarding the acceleration efficiency of the candidate accelerator and may also provide an estimate of the energetic particle flux precipitating on the solar surface. 
For this purpose we would need to know the properties of the shock interaction with the magnetic fields going and those returning to the Sun. Because the solar corona is not yet accessible to in situ measurements, we rely here on remote sensing and high-energy radiation diagnostics to infer the conditions in which particles are accelerated.

In this study we concentrate on deriving the shock properties of three far-side CME events that were associated with $\gamma$-ray emissions with very different intensities and durations. Section~\ref{sect:obs} presents an observational overview and analysis of the three events measured as radio, visible, extreme ultraviolet (EUV), and high-energy X-ray and $\gamma$-ray radiation. The measurements of energetic particles in situ by distant spacecraft are reviewed in section~\ref{sect:sep_obs} and in appendix~\ref{sec:sep}. We use a combination of reconstruction techniques and numerical models to derive the time-dependent properties of the expanding shock wave (velocity, Mach number and geometry) in section~\ref{sect:models}. In section~\ref{sect:evolution} we test the idea that long-duration $\gamma-$ray emissions result from energetic particles produced by the coronal shock region that is magnetically connected with the visible solar disk.  We discuss the main findings of the study and draw conclusions in section~\ref{sect:discussion}.

\section{Observations} \label{sect:obs}

The three far-side eruptions on 2013 Oct 11, 2014 Jan 06, and 2014 Sept 01 were observed by a number of ground- and space-based observatories.  We first discuss the observations of the flares and the time-extended $\gamma$-ray measurements.  In the next section we discuss the CME observations, the evolution of the shock front, and magnetic connections to spacecraft that measure the SEPs discussed in the last section.  

\subsection{Overview of flares and $\gamma$-ray observations}
% In this first part we briefly review the initial observations of the flares that triggered the CME shocks and sustained high-energy X-ray/$\gamma$-ray emission. \\
\indent In Figures~\ref{fig:gammaObs_1}-\ref{fig:gammaObs_3}  we plot the `light-bucket' flux time histories of $>$100 MeV $\gamma$-rays observed by LAT in the three events discussed in this paper. In the main figure, we plot the fluxes within 10 degrees of the Sun in the hours around the flare.  Discrete LAT observing intervals with significant solar exposure are typically about 20-40 min in duration and occur every other orbit.  The background fluxes are predominately comprised of Galactic, extragalactic, and quiescent solar photons. Details of these and 26 other sustained-emission
$\gamma$-ray events observed by Fermi are discussed in (Share et al. 2017, to be submitted).

\subsubsection{2013 Oct 11}
\label{subsect:flare_oct13}
On 2013 Oct 11 at 07:01 UT a flare occurred from the active region (AR) located at N21E106 position and was observed in soft X-rays by the Geostationary Operational Environmental Satellite Network ({\it GOES}) and the MErcury Surface, Space ENvironment, GEochemistry and Ranging ({\it MESSENGER}). In addition, a wealth of  EUV, visible, radio, X-ray and $\gamma$-ray radiations were observed by the the Solar-Terrestrial Relations Observatory \citep[STEREO;][]{2008SSRv..136....5K}, the Solar and Heliospheric Observatory ({\it SoHO}) and the Solar Dynamics Observatory \citep[{\it SDO};][]{2012SoPh..275...17L}, {\it RHESSI}, {\it GOES,} and Fermi spacecraft \citep[see, e.g.,][]{2015ApJ...805L..15P, 2017ApJ...835..219A}. The AR was 10 degrees behind the east limb viewed from Earth at the time of the flare. A drifting metric radio type II was measured between 07:10 and 07:20 UT by the Culgoora spectrograph with a drift speed of 924 km/s, close to the derived CME/shock speeds measured in Figure~\ref{fig:bulle_apex}(d) during that time interval, as shown later in this work.

\begin{figure}
   \centering
  \includegraphics[width=0.48\textwidth]{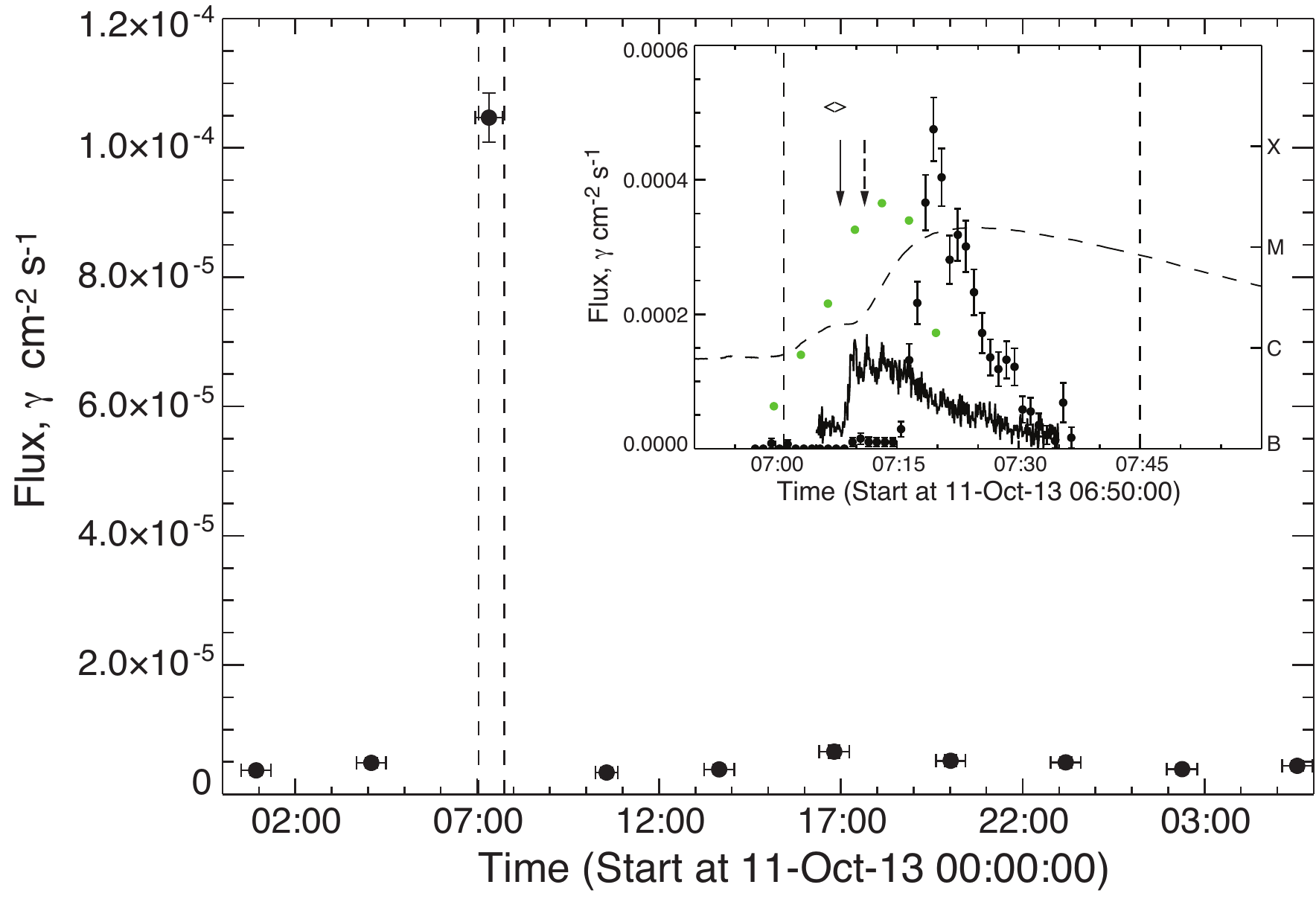}
\caption{Time history of the $>$100 MeV flux from $\leq$10 degrees of the Sun revealing the 2013 Oct 11 LAT event.  Vertical dashed lines show the {\it GOES} start and end times. The inset shows 1 min accumulation LAT flux data points and arbitrarily scaled 50-100 keV count rates observed by {\it Fermi }/GBM. The dashed curve shows the {\it GOES} 1 -- 8{\AA} profile (scale on right ordinate) and the $<->$ symbol the estimated time of the CME onset from the CDAW catalog.  The vertical solid arrow depicts our estimate of the CME onset from inspection of {\it SDO}/AIA images; the type II radio emission began about 2 min later. The derivative of the 1--4 keV soft X-ray emission observed by Solar Assembly for X-rays (SAX) instrument on {\it MESSENGER} \citep{schl07} that directly observed the flare is shown by the green dots. This reflects the hard X-ray profile of the flare.}
              \label{fig:gammaObs_1}%
 \end{figure}

As the flare was located beyond the east limb of the Sun, the soft X-ray emission recorded by {\it GOES} came from the upper corona.  The emission from this vantage point lasted $\sim$44 min and is shown by the dashed vertical lines in Figure~\ref{fig:gammaObs_1} and by the dashed curve in its inset.   The Solar Assembly for X-rays (SAX) instrument on {\it MESSENGER} \citep{schl07} directly observed soft X-rays from the flare site (red dots) and showed that the 1--4 keV emission from the flare peaked more rapidly than that observed by {\it GOES}.   The onset of the CME occurred about 7 min after the start of the soft X-ray emission and about 3 min before type ll metric radio emission was detected.  The green dots represent the derivative of the 1--4 keV emission observed by SAX and provide a surrogate for the time profile of hard X-ray emission from the flare site.  Hard X-ray emission from the flare began $\sim$ 07:00 UT and peaked $\sim$ 7 min before the peak in soft X-rays.  {\it Fermi}/GBM observed a precipitous rise in 50--100 keV X-ray emission about 07:08 UT, which appears to reflect the time that flare hard X-rays in the corona became visible above the limb of the Sun, as viewed from Earth. The{ \it RHESSI} hard X-ray images up to 50 keV indicate that the source of the emission was above the limb of the Sun \citep[][]{2015ApJ...805L..15P}.  We fit the background-subtracted hard X-ray spectrum observed from 15--200 keV between 07:09 and 07:12 UT;  the emission appears to extend up to $\sim$100 keV and can be fit equally well by a spectrum of electrons with power-law index of --4.8, interacting in a thick target, or a spectrum with a power-law index of --3, interacting in a thin target.  

The LAT had good solar exposures every other orbit on 2013 Oct 11. The exposure from 06:58--0740 UT captured all of the time in which hard X-ray emission was observed. As shown in the figure, emission $>$100 MeV was only observed during this first orbit. There was no evidence for emission in the next exposure three hours later between 10:16--10:50 UT.  The source-class data plotted at 1 min resolution in the inset reveal an increase in $>$100 MeV emission beginning at 07:15 UT that peaks in 5 min and falls back to background by 07:40 UT. The centroid of the $>$100 MeV emission by LAT is consistent with a location near the east limb of the Sun at N03E62 with a 1$\sigma$ range in longitude from E39 to just above the eastern limb \citep{2015ApJ...805L..15P}.  The background-subtracted $\gamma$-ray spectrum from 07:14--07:30 UT is consistent with emission from pion decay produced by a proton spectrum $>$300 MeV, following a differential power law with an index --3.7 $\pm$ 0.2 with no evidence for spectral variation during the rising and falling phases.  A better fit might be achieved with a proton spectrum rolling over at energies above 500 MeV.   These measurements indicate that the protons producing the $\gamma$-ray emission interacted deep in the solar atmosphere on the visible disk as viewed from Earth.

\subsubsection{2014 Jan 06 event}
\label{subsect:flare_jan14}
\begin{figure}
   \centering
  \includegraphics[width=0.48\textwidth]{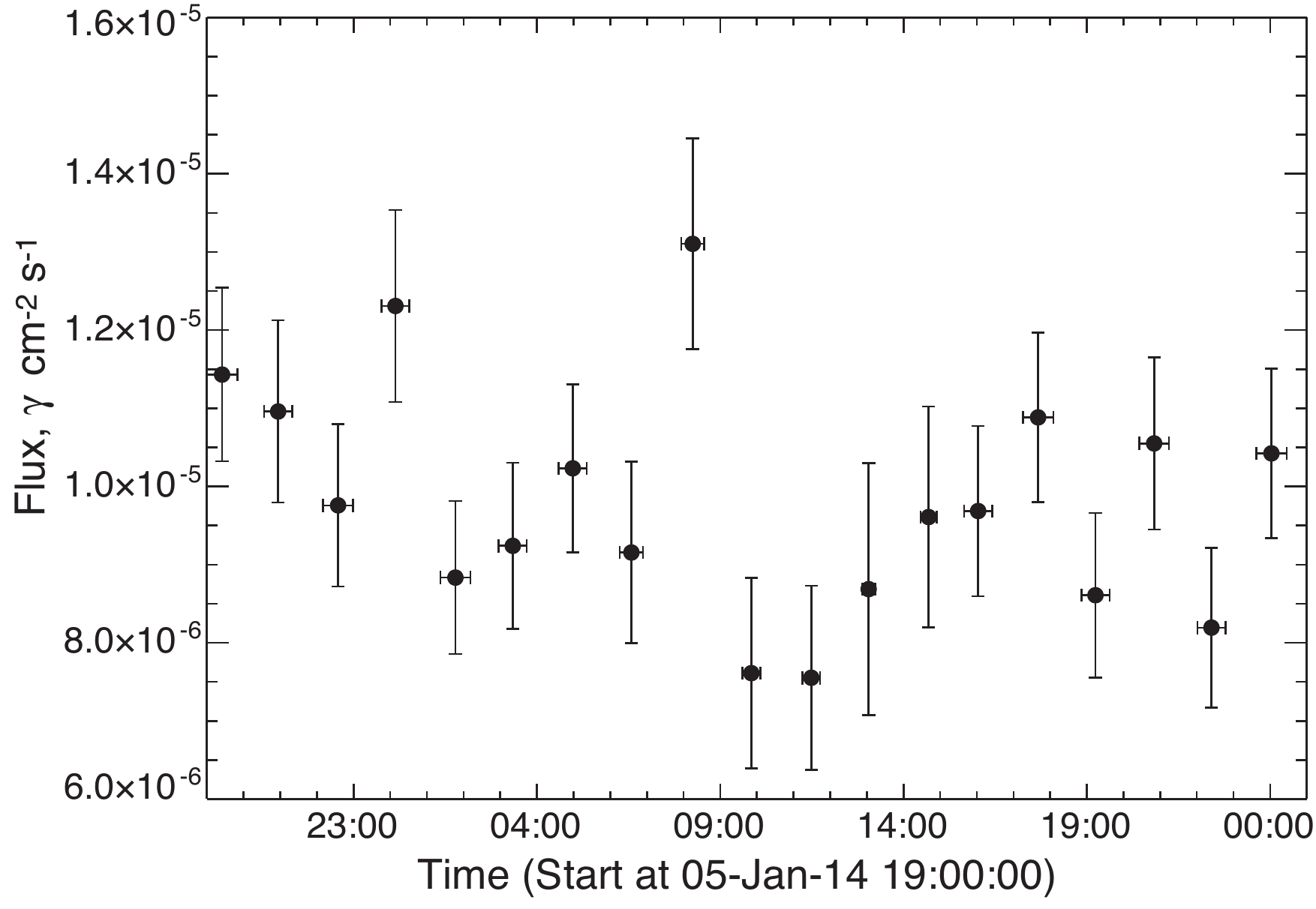}
\caption{Time history of the $>$100 MeV flux from $\leq$10$^{\circ}$ of the Sun around the time of the reported $>$100 MeV event on 2014 Jan 6.}
              \label{fig:gammaObs_2}%
 \end{figure}

On 2014 Jan 06 a C2.2 flare occurred at about 07:30 UT \citep{Thakur14} from an AR located at S8W110 and a large filament eruption was observed at 07:44 UT. The {\it GOES} soft X-ray flare class does not represent its real magnitude because the AR was $\sim20$ degrees behind the solar limb. Metric type II emission was observed between 07:45 and 08:05 UT by the Learmonth spectrograph, but no speed could be derived using the spectrograph. Type III radio emission started at 07:45 UT as seen by wave and S-wave receivers on board both {\it STEREO} and {\it Wind} spacecraft. The $\gamma$-ray emission from this behind-the-limb flare was reported by \citet{2015arXiv150704303P, 2017ApJ...835..219A}.  However, as can be seen in Figure~\ref{fig:gammaObs_2}, this excess between 07:55 and 08:30 UT  was not significant using our light bucket analysis and did not meet our criteria for detection above background.  We find no evidence for variation in the $>$100 MeV flux in that time. This solar event was associated with energetic particles measured near Earth and deserves a thorough analysis in order to understand why the $\gamma$-ray fluxes were so weak.

\subsubsection{2014 Sept 01 event}
\label{subsect:flare_sept14}
On 2014 Sept 01 a solar flare occurred at about 10:54UT and lasted 40 minutes as deduced from the on-disk observations made by the MESSENGER/SAX soft X-ray monitor. The AR was located at N14E126 corresponding to 36 degrees behind the east limb viewed from Earth. Metric emission was observed with a drift speed of 2079 km/s by the Nan\c cay  {\it ORFEES} spectrograph between 11:00 and 11:50 UT \citep{2015arXiv150704303P}. In addition the Nan\c cay radioheliograph imaged the type II burst from 11:00 UT in a region situated near the forming CME. A type III burst was detected at 11:02$\pm$00:02 UT by {\it STEREO} and  {\it Wind} radio and plasma wave receivers.

\begin{figure}
   \centering
  \includegraphics[width=0.48\textwidth]{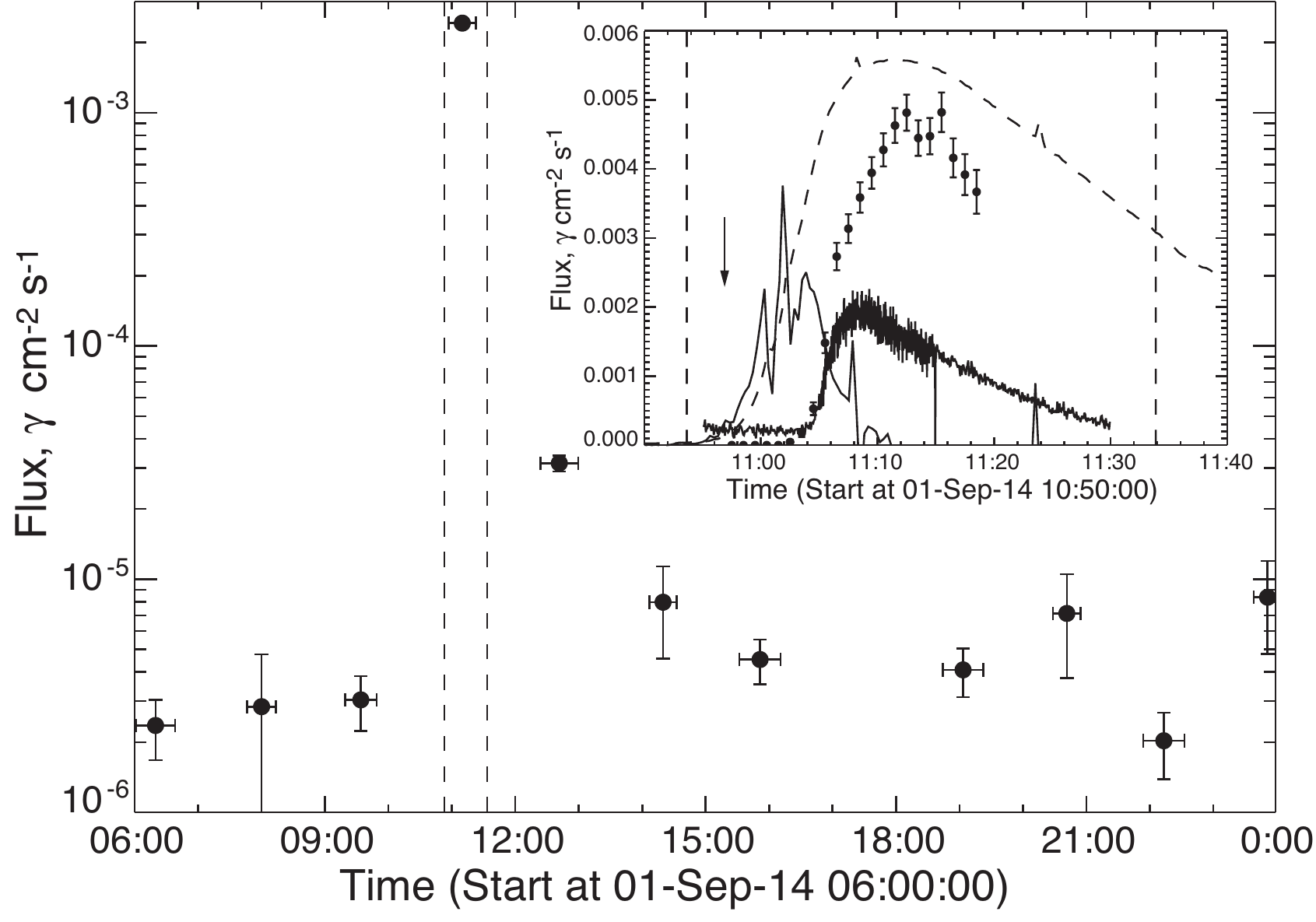}
\caption{Time history of the $>$100 MeV flux from $\leq$10$^{\circ}$ of the Sun revealing the 2014 Sept 1 LAT event.  Vertical dashed lines show the estimated soft X-ray start and end times from the SAX instrument on {\it MESSENGER} that directly observed the flare. The inset shows 1-min accumulation LAT flux data points and arbitrarily scaled 100-300 keV count rates observed by {\it Fermi }/GBM. The dashed curve shows the 1--4 keV time profile observed by the SAX instrument.  The vertical solid arrow depicts our estimate of the CME onset from inspection of {\it SDO}/AIA images and the type ll radio emission began about 3 min later.  The derivative of the 1--4 keV emission plotted as the thin solid curve is an estimate of the hard X-ray time profile at the flare site.}
              \label{fig:gammaObs_3}%
 \end{figure}

The flare associated with this event occurred behind the solar limb at about E126.  At this location the coronal region above the flare was not visible at Earth and no soft X-ray emission was observed by {\it GOES}.  However, the flare was observed by the Solar Assembly for X-rays (SAX) instrument on {\it MESSENGER} \citep{schl07}.  The soft X-ray time profile observed by SAX is plotted as the dashed curve in the inset of Figure~\ref{fig:gammaObs_3}; this emission had a duration of about 40 min.  From {\it SDO} 193{\AA}, 211{\AA} images, we estimate that the onset of the associated CME occurred about 10:57 UT, at the rise of the inferred flare hard X-ray emission. The latter was computed by taking the derivative of the SAX soft X-ray time profile plotted as a solid curve in the inset.  

{\it Fermi}/LAT had good solar exposures every other orbit and exposures about four times smaller in the intervening orbits.  Such a 25\% exposure occurred during the behind-the-limb flare from about 11:06--11:20 UT when the Sun was at a large angle with respect the LAT telescope axis.  Source-class LAT data could be used to study $>$100 MeV $\gamma$-ray emission because the intense hard X-ray emission from the flare did not reach {\it Fermi}.  At these large solar viewing angles, the detector response is small and not accurately determined.  There were also two good exposures 12:26--12:58 and 15:36--16:08 UT during which LAT had significantly higher sensitivity to search for delayed high-energy emission.  The $>$100 MeV flux plotted in the main figure just after the flare was one of the largest observed by LAT during its eight-year mission.  The hourly fluxes are plotted logarithmically to reveal the sustained emission that lasted about 3 hours. In the inset we plot $>$100 MeV fluxes (individual data points) derived from the Pass 8 source-class data at 1-min resolution along with arbitrary 100-300 keV rates (solid data curve) observed by GBM.  Both the $\gamma$-ray and 100--300 keV X-ray emissions, as viewed from Earth, appear to rise at $\sim$11:04 UT, about 7 min after the onset of the hard X-ray emission observed from the flare as viewed by {\it MESSENGER}.  The hard X-rays observed by GBM peaked by 11:08 UT and fell exponentially while the $>$100 MeV $\gamma$-ray flux peaked about 5 min later and appeared to fall as rapidly as it increased.  The sustained $\gamma$-ray emission observed between 12:26 and 12:58 UT is likely to be the tail of this earlier post-impulsive phase emission.

We have fit the background-subtracted $>$100 MeV $\gamma$-ray data with a pion-decay spectrum produced by $>$300 MeV protons following a differential power-law spectrum and interacting in a thick target at a heliocentric angle of 85$^{\circ}$.  Our fits suggest that the spectrum hardened during the duration of the event with spectral indices of --4.25 $\pm$ 0.15, --3.85 $\pm$ 0.1, and --3.45 $\pm$ 0.35, at 11:06--11:12, 11:12--11:20, 12:26--12:58 UT, respectively.  We searched for the presence of the 2.223 MeV neutron capture line in the GBM BGO spectrum between 11:04 and 11:30 UT and set a 95\% confidence upper limit of 0.016 $\gamma$ cm$^{-2}$ s$^{-1}$ on its flux.  Comparing the $>$100 MeV $\gamma$-ray fluence with the upper limit on the 2.223 MeV line fluence, we estimate that the proton power-law spectral index between 30 and 300 MeV was harder than --3.4; this was estimated assuming that the protons interacted at a heliocentric angle of 85$^{\circ}$.  The index would be even harder for
smaller heliocentric angles. This suggests that the spectrum of protons producing the pion-decay emission steepened above 300 MeV.

The NaI detectors on GBM observed a hard X-ray spectrum up to at least 1 MeV.  Our fits to the 20--900 keV spectrum  between 11:06 and 11:16 UT indicate that the hard X-rays were produced by a power-law electron spectrum with index --3.1 and low-energy cutoff of 130 keV for a thick target model with a 5\% confidence in the quality of the fit.  With this high low-energy cutoff in the electron energy we would not expect to observe significant soft X-ray emission from chromospheric evaporation on the visible disk and such emission was not detected by {\it GOES}. The fit to the spectrum for a thin target model is significantly worse (confidence $< 10^{-6}$ \%) than for a thick target.  There is no evidence for variation of the electron spectrum over the event. We also fit the GBM BGO spectrum up to 40 MeV with an electron spectrum having a power-law index  of --3.1 interacting in a thick target. Thus the results indicate that the bremsstrahlung originated from a single spectrum of electrons with a high-low energy cutoff and power-law index of $\sim$3.1 up to tens of MeV interacting in a thick target.

The fact that both the protons producing the $\gamma$-ray emission and the electrons producing bremsstrahlung up to about 40 MeV interacted in a thick target on the visible disk and that both emissions commenced at the same time, suggests that they had a common origin. In the next sections, we investigate whether this common origin could be the CME shock. 

      \begin{table*}
                \caption[]{Sequence of first detection times by different instruments for the three studied events. First observation times in radio type II and III, soft and hard X-ray, and high-energy $\gamma$-ray observations are presented. Energetic particle (high-energy protons and electrons at 1 AU and at Mars) onset times are given as well. }
      $$
         \begin{array}{p{0.3\linewidth} l c  c  c }
            \hline
            \noalign{\smallskip}
             Waveband or Instrument  & {\rm 2013~Oct~11}  &  {\rm  2014~06~Jan} &   {\rm  2014~Sep~01}\\
            \noalign{\smallskip}
            \hline
            \noalign{\smallskip}
            \underline{\it EM emission:}   &   & &  \\
            {\rm Radio type II} & 07:10 & 07:45 & 11:00  \\
            {\rm Radio type III}  & 07:15 & 07:45 & 11:00 \\
            {\rm Soft X-ray}  & 07:01  & 07:42 &  10:54    \\
            {\rm Hard X-ray}  & 06:58  &  &   10:57   \\
            {\rm Fermi LAT $\gamma$-ray ($>100$ MeV)}  & 07:15  & Too~Weak &   11:04   \\
            \noalign{\smallskip}
            \hline
            \noalign{\smallskip}
            {\rm Shock connectivity to vis. disk}  & 07:10  & 07:45 &   11:05   \\
             \noalign{\smallskip}
            \hline
            \noalign{\smallskip}
            \underline{SEP~electrons (1~AU):}  &   &  &     \\
             {\rm {\it STEREO}-A; HET~(0.7-4~MeV)}  & 07:29 & 09:30  &   11:11  \\
             {\it STEREO}-B; HET~(0.7-4~MeV)  & 07:40  & Too~weak &   11:25   \\
             L1; EPHIN (0.25-10~MeV)  & 13:00 \pm 30  & Data~gap &   14:22   \\
             \noalign{\smallskip}
            \hline
            \noalign{\smallskip}
             \underline{SEP~protons (1~AU):}  &   &  &     \\
             {\rm {\it STEREO}-A; HET~(29.5-100~MeV)}  & 07:58 & 09:30  &   Data~gap  \\
             {\it STEREO}-B; HET~(29.5-100~MeV)  & 08:19  & 18:45 &   11:48   \\
             L1; EPACT (28-72~MeV)  & 22:30~(+1~day)  & 08:25 &   00:15~(+1~day)   \\
             GOES 13 ($>100$~MeV)  &   &  &   18:00 \pm 01:00   \\
             \noalign{\smallskip}
            \hline
            \noalign{\smallskip}
             \underline{SEP on Mars:}  & 07:45  & 09:00  &  12:00  \\
            \noalign{\smallskip}
            \hline
         \end{array}
    $$
          \label{table2}
   \end{table*}

\subsection{Solar energetic particles measured in situ}
\label{sect:sep_obs}
Strong solar energetic particles fluxes were also measured in situ at widely separated vantage points during the three $\gamma$-ray events by {\it STEREO} and by spacecraft situated along the Sun-Earth line, i.e., {\it SoHO}, {\it Wind}, and the Advanced Composition Explorer ({\it ACE}), and at Mars. Particle fluxes measurements provide relevant diagnostics of the particle acceleration in the low corona, which are exploited later in this article to reach a comprehensive interpretation of the SGRE and SEPs in terms of the expanding coronal shock. The in situ measurements are highly complementary for the following reasons.

\begin{itemize}
\item The onset of the relativistic electron flux and its intensity measured in situ at the spacecraft, along with their magnetic connectivity to the Sun, provide a direct measurement of the level of energization of particles along a given magnetic field line and information on the energetic electron solar release times.
\item The peak level of the particle flux and its integrated value over the event provides information on the number of accelerated particles that propagate outward from the Sun along specific field lines. With certain assumptions, particle flux can also be used to estimate the number and energy of particles propagating sunwards \citep[e.g.,][]{Cliver93},
\item Solar energetic particle spectra can be compared with the particle spectra at the Sun derived from hard X-ray $\gamma$-ray emission, although the in situ spectra are likely modified by particle propagation and diffusion in the heliosphere.
\item When measurements from several well-separated spacecraft are available, the rapid increase of SEP flux at these probes provides information on particles accelerated across a broad range of heliographic longitudes \citep{1996ApJ...466..473R, 1998ApJ...509..415L, 2012SoPh..281..281D, 2012ApJ...752...44R, 2014ApJ...797....8L}.
\end{itemize}

Several studies have jointly discussed the sustained $\gamma$-ray emission and SEPs \citep[e.g.,][]{1987ApJ...316L..41R, Ryan00, 2009RAA.....9...11C,2017ApJ...835..219A}. Some recent works studied jointly SEPs and the expanding coronal shock with similar methods as developed in the present study. Because the main focus of this paper is the relation between the SGRE and expansion of coronal shocks, we placed our analysis of the SEPs measured in situ in the appendix. The conclusions drawn from this analysis are that, all spacecraft considered, the strongest measured SEPs were during the 2014 Sep 01. The peak SEP flux measured on 2013 Oct 11 and  2014 Jan 06 were one order and three orders of magnitude lower than the 2014 Sep 01 SEP event, respectively. This hierarchy follows that of the SGRE peak intensities for the three events. Probes well connected to the CME source region measured a significant increase of the highest SEP flux between 10 MeV and 100 MeV.  Probes connected to the solar surface far from the CME source longitude measured much lower levels of SEP flux. The STEREO electron flux was 50-100 times higher during the 2014 Sept 01 event than during the 2013 Oct 11 event, we use this observation in the discussion section to interpret the relative strength of any associated sustained electron emission as inferred from $>$100 keV bremsstrahlung. As shown in the following,  the in situ measurements suggest that a common accelerator is acting in the corona to produce the $\gamma$-ray and the SEP events. The onset times of the particle flux increases measured the in situ data are included in Table~\ref{table2}.

\subsection{Coronal mass ejection propagation and 3D determination of the pressure wave fronts}
\begin{figure}
   \centering
  \includegraphics[width=0.48\textwidth]{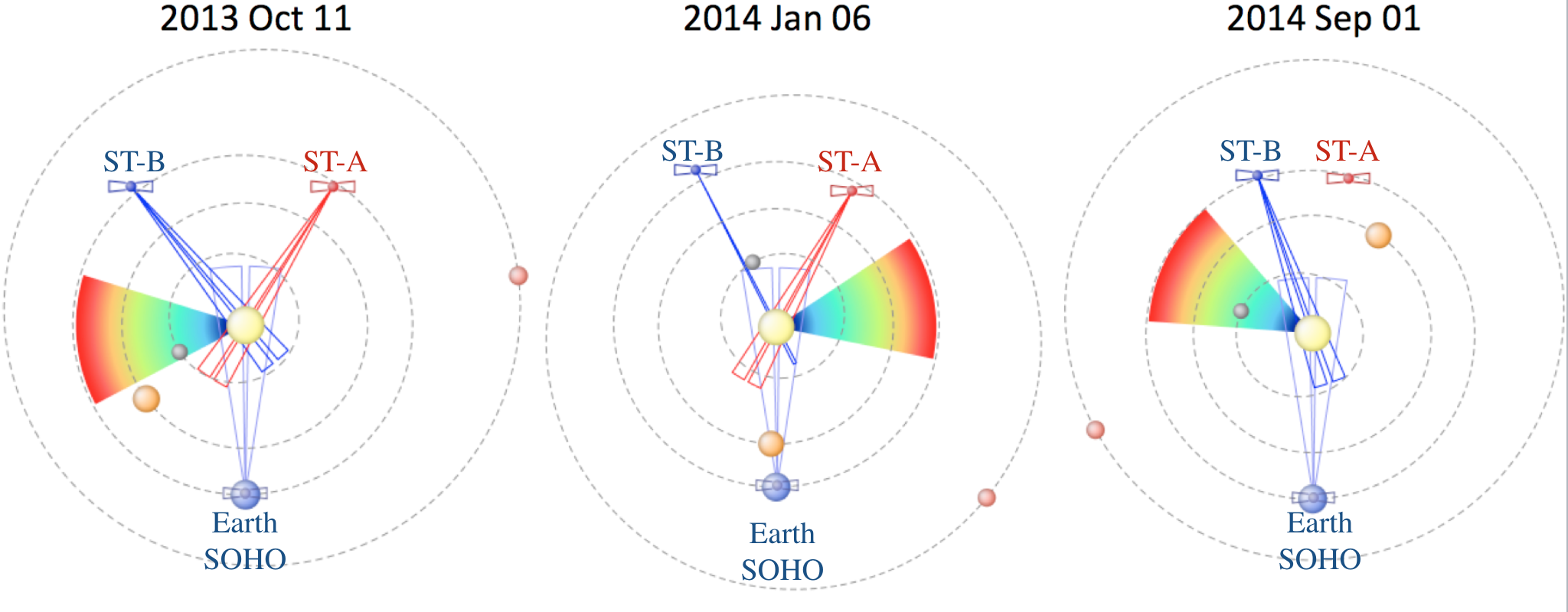}
   \caption{View of the ecliptic plane from solar north of the orbital configuration of the four inner planets (plotted as spheres), the {\it SoHO} spacecraft, and the  {\it STEREO} spacecraft (symbols). The Earth position is fixed to the bottom of the figures. The extent of the camera field of view in the ecliptic plane used in the triangulation of the  CME pressure wave are plotted for each case. The direction of propagation of the CME is shown as a color-coded cone of fixed longitudinal extent. These directions of propagations were derived by the HELCATS project using heliospheric imaging (except for the 2014 Sep 01 case, in which the coronal shock nose propagation direction was used).}
              \label{fig:orbital_conf}%
\end{figure}

    In our study of the three CMEs associated with the behind-the-limb events we used complementary data sets from {\it STEREO}, {\it SoHO,} and {\it SDO}. For each event, the CME eruption and propagation were observed in EUV ({\it SDO} and {\it STEREO}) and in white-light images ({\it {\it SoHO}} and {\it STEREO}) from observatories situated at least two of the three available vantage points.
    
 Figure~\ref{fig:orbital_conf} shows the positions of the four observatories used to triangulate the CME extent in three dimensions. The {\it SDO} location is the same as Earth and {\it SoHO} on these scales. Also shown are the fields of view of the instruments and the locations of the four inner planets. A first rough estimate of the directions of propagation for the 2013 Oct 11 and 2014 Jan 6 events was deduced from solar wind imaging and given in the catalog of CMEs produced by the Heliospheric Cataloguing, Analysis and Techniques Service (HELCATS) project\footnote{Project website: https://www.helcats-fp7.eu/}. As there was no solar wind imaging available for the 2014 Sep 01 event, the direction of propagation of this CME is defined as the nose of the triangulated shock (derived in the next paragraphs). The heliocentric longitudes of propagation of the 2013 Oct 11 and 2014 Sep 01 eruptions were behind the eastern solar limb while that of the 2014 Jan 06 eruption was behind the western solar limb.

To derive the 3D extent of the CME low in the corona we used EUV observations from ({\it SDO} and {\it STEREO}) and white-light observations from ({\it SoHO} and {\it STEREO}). Our requirement was that imaging data was acquired from at least two of the three available vantage points. The 3D triangulation technique employed is described in detail in \citet{Rouillard16}. In a nutshell, the technique tracks in 3D the region perturbed by the pressure front generated by the expansion of the CME. This region usually consists of a bubble visible off-limb combined EUV-white-light images as well as a wave front visible on disk in EUV images \citep{2009AIPC.1183..139V,2009ApJ...700L.182P}. In this reconstruction technique the wave front is considered to be the intersection between the 3D bubble, modeled as an ellipsoid, and the coronal region located just above solar surface imaged in EUV. The bubble-EUV wave system is considered to be the outermost extent of the region perturbed by the CME as it forms and expands in the corona. Areas of the bubble surface moving at speeds exceeding the local characteristic speed of the coronal medium (through which the bubble is propagating) are the location where a shock wave is likely to form.

\begin{figure*}
   \centering
  \includegraphics[width=0.85\textwidth]{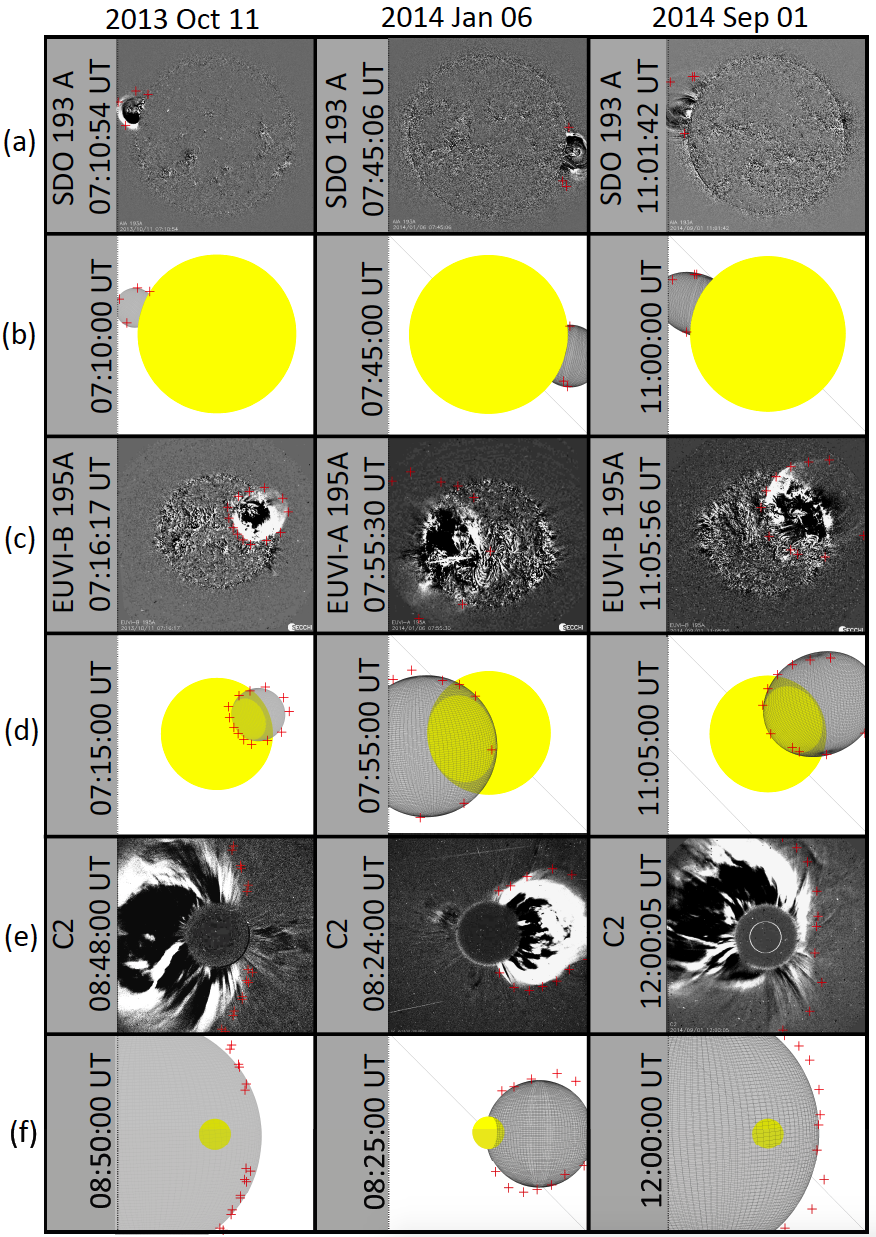}
   \caption{Comparison of running-difference images (rows a,c,e) of the three events with the results of applying the fitting technique (rows b,d,f) developed in \citet{Rouillard16}. The images are from the {\it SDO} AIA and the {\it STEREO} EUVI instruments (rows a and c) and from the {\it STEREO}-LASCO C2 instrument (row e). Red crosses superimposed on the fitted ellipsoids show the measured contours of the propagating fronts observed in the running difference images and are used to constrain the extent and location of the fitted ellipsoids at each time.}
              \label{fig:TRIANG}%
 \end{figure*}

 We show a selection of images for the three events from the three instruments in Figure~\ref{fig:TRIANG}a,c,e. We manually extracted the location of the EUV wave and the CME bubble at all available times and from all available viewpoints (EUVI and coronagraphs on board {\it SDO}, {\it SoHO,} and {\it STEREO} spacecraft). We used EUV and white-light images from the three vantage points for the 2013 Oct 11 and 2014 Jan events but only from two vantage points for the 2014 Sept 01 event because no {\it STEREO} A data was available. The manually selected points are plotted as red crosses on the images of the CMEs. The CME events erupted on the far side of the Sun and it was essential that the {\it STEREO} EUV instruments provided on-disk images of the source regions situated on the far side. This allowed us to track the expansion of the CMEs during the early stages of their formation; in particular, we tracked the EUV wave expansion.  The manually selected points plotted as red crosses on the images are also plotted on the cartoons of the CMEs in rows (b), (d), and (f) of Figure~\ref{fig:TRIANG}. We assumed that a 3D ellipsoid could model the pressure front and chose the dimensions of the ellipsoid (defined by a set of its three axis values and its central position) such that the locations of the red crosses, as viewed from the different vantage points, best reflected the rendered ellipsoid. This included the two constraints that the ellipsoid, as viewed off the solar limb passed by the outline of the bubble and that the  intersection of the ellipsoid with the solar surface passed by the EUV front. We obtained a set of fitted ellipsoids covering the first hour of the CME expansion as viewed from Earth and from at least one of the {\it STEREO} spacecraft.

 We follow the method outlined in \citet{Rouillard16} to compute the 3D expansion speed of the shock surface. This procedure first involves selecting a grid of points on the fitted ellipsoid at time $t$ and then finding the locations of the closest respective points on the ellipsoid at previous time-step $t-\delta t$. The distance traveled between these points was divided by the time interval $\delta t$ to obtain an estimate of the 3D speed of the shock surface as a function of time. We list the maximum CME speeds for the three events in Table~\ref{table1}. \\

   \begin{table}
             \caption[]{Table relating the dates of eruptions, CME pressure wave maximal speeds as derived by the triangulation technique, and solar wind speed values measured in situ at the three vantage points at 1~AU (Lagrange point L1, STEREO-A and STEREO-B)}
      $$
         \begin{array}{p{0.2\linewidth} l c  c}
            \hline
            \noalign{\smallskip}
           Date    & V_{\rm CME, max}  & {\rm V_{sw}}  \\
               & {\rm (km/s)} &  {\rm (km/s)}  \\
                       &  &  {\rm L1~|~STA~|~STB}  \\
            \noalign{\smallskip}
            \hline
            \noalign{\smallskip}
            2013 Oct 11  &  1580   & 394~|~343~|~323 \\
            2014 Jan 06 &1880  & 396~|~351~|~383 \\
            2014 Sep 01  &  2630  & 443~|~418~|~444      \\
            \noalign{\smallskip}
            \hline
         \end{array}
    $$
          \label{table1}
   \end{table}

      \begin{figure*}
   \centering
  \includegraphics[width=0.98\textwidth]{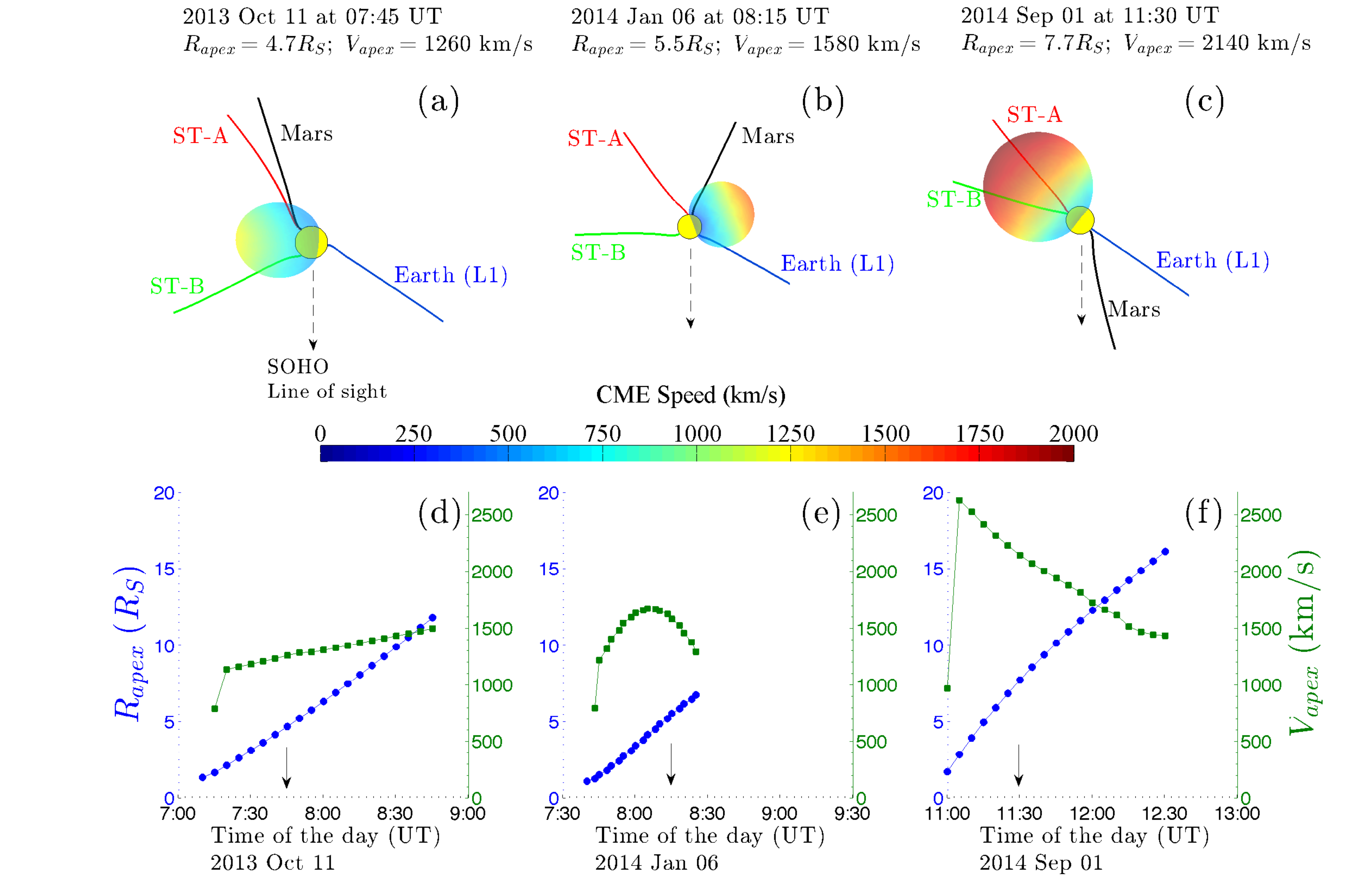}
   \caption{Top panels a, b, and c: Three views of the ecliptic plane from solar north depicting the ellipsoids of the three CME relative to the Sun (drawn as a yellow filled circle) approximatively 30 minutes after the accompanying type II radio bursts. The downward arrows depict the Earth's location. The derived 3D shock speeds determined by the triangulation technique are shown by the coded colors superimposed on the ellipsoids. The four colored lines depict the magnetic field lines, which are connected to the spacecraft making in situ measurements. Bottom panels d, e, and f: The time variation of the derived speed and heliocentric distance of the apex of the CME front for the three CME events. The black vertical arrows indicate the times corresponding to the snapshots in the top panels.}
              \label{fig:bulle_apex}%
    \end{figure*}

The results of our analysis are provided in Figure~\ref{fig:bulle_apex} for the three events. In panels (a-c) we show the extent of the pressure waves in the ecliptic plane 30 minutes after the onset of radio type III bursts.  The speeds at different points of the shock-front ellipsoids are color coded.  The maximum speed at any one time is typically found at the nose (apex) of the structure.  For example, the nose of the 2014 Sep 01 event reached a speed of  2140 km/s and is by far the fastest event of the three considered in this study. The maximum speeds reached at any time during the three CME expansions are given Table~\ref{table1}.  In panels (d-f) we plot the time evolution of the radial extent and the maximal speed of the shock front for each event. 
The 2013 Oct 11 event accelerated rapidly in the low corona from 07:10 to 07:20~UT up to 1100 km/s and its corresponding height was $2 R_\odot$. Then CME continued its acceleration phase much more gradually, reaching 1500 km/s by 08:45~UT and corresponding height of $\simeq12 R_\odot$. The 2014 Jan 06 pressure wave was fitted over a 1 hour time interval. Its initial acceleration phase started at $\sim$07:40~UT and continued up to 08:10~UT when the maximum speed was reached, 1880 km/s, and corresponding apex height of $\simeq5 R_\odot$. After this phase the pressure wave started decelerating. The acceleration phase was more abrupt and shorter lived during the 2014 Sep 01 event; it reached 2630 km/s between 10:55 to 11:05. The deceleration phase started just after and the front speed was 1500 km/s when the apex height was  $\sim17 R_\odot$ (by 12:30~UT). These different expansion profiles are likely the result of different properties of the magnetic piston driving the eruption of the CME.

Also shown in panels (a-c) of Figure~\ref{fig:bulle_apex} are representative magnetic field lines connecting the different spacecraft considered in this study and the expanding pressure front.  We followed the procedure outlined in \citet{2014ApJ...797....8L} and \citet{Rouillard16}, which assumes the interplanetary medium consists of Parker spirals defined by the solar wind speeds measured at the connecting spacecraft; these speeds are given in the Table~\ref{table1}. This spiral is traced from the spacecraft down to 15 R$_\odot$ and the coronal magnetic field is obtained by a 3D MHD model.  This  Magneto-Hydrodynamic Around a Sphere Thermodynamic (MAST) model is described in $\S$~\ref{subsect:mast}  (Lionello et al. 2009). This mapping of magnetic connectivity is exploited to discuss the complementary SEP measurements acquired in situ during the three SGRE events analyzed here.

  % \section{Shock properties and magnetic connectivity using coronal field models}
  %  \section{Magnetic conductivity of particles accelerated at the shock front to the Sun and space}
  % ... particles have no connectivity by themselves. Also we do not study conductivity properties, in strict sense of the term.
        \section{Magnetic connectivity of the shock front to the Sun and space}

   \label{sect:models}
   
 We developed a model to test whether particles accelerated at the shock front of a CME can explain the sustained $>$100 MeV $\gamma$-ray emission observed by Fermi.  This is based on our ellipsoid approximation for the pressure wave and knowledge of the coronal magnetic field lines that this pressure wave intersects as a function time, some of which can reach the visible disk of the Sun as view from Earth. We derived the magnetic connectivity lines using two different methods in the next two sections. In the last section, for the three flares, we project the velocities of the shock as it crosses these field lines onto the locations where they intersect the solar photosphere at different times in its expansion.  A similar process allowed us to project these velocities onto field lines reaching the spacecraft making in situ particle measurements.    

\subsection{The Potential Field Source Surface (PFSS) model}
 \label{subsect:pfss}
The PFSS reconstruction technique \citep[e.g.,][]{1992ApJ...392..310W} can be used to model the coronal magnetic field between one $R_{\odot}$ and a source surface, typically placed at 2.5 $R_{\odot}$. This model only requires a magnetogram of the Sun on a given date. The two main assumptions of this technique are the absence of currents in the corona and the presence of a spherically uniform source surface. The latter assumption forces a rapid expansion of the magnetic flux tubes between the solar surface and source surface, where the field lines are forced to be radial. While this model provides fairly accurate estimates of the global magnetostatic topology of the corona, it does not give information about other required plasma quantities, such as local density and temperature and therefore cannot be used alone to provide the Mach number of the shock. The version of the PFSS model used here comes from the Lockheed Martin Solar And Astrophysics Laboratory (LMSAL)\footnote {http://www.lmsal.com/~derosa/pfsspack/}. The field extrapolation is based on evolving surface magnetic maps constructed from data taken by the Helioseismic and Magnetic Imager \emph{Helioseismic and Magnetic Imaging}  \citep[HMI;][]{2012SoPh..275..207S} onboard {\it SDO}. These maps account for the transport and dispersal of magnetic flux across the photospheric surface using a flux-transport model \citep{2003SoPh..212..165S}. The magnetic transport processes are differential rotation and supergranular diffusion and they continuously modify the distribution of photospheric magnetic fields. In the analysis that follows, we assume that the magnetic field is purely radial above the source surface. The magnetic flux redistribution  can result in highly nonradial field lines in between the solar surface and source surface at 2.5 $R_{\odot}$. At this external boundary the magnetic field is constrained to be purely radial. In this study, we used the PFSS model primarily to compare the derivation of magnetic connectivity obtained with magnetohydrodynamic models described in the next section. 
    
\subsection{MAST model}
\label{subsect:mast}
Magneto-Hydrodynamic Around a Sphere Thermodynamic (MAST) model is a MHD model \citep{2009ApJ...690..902L} that includes detailed thermodynamics with realistic energy equations accounting for thermal conduction parallel to the magnetic field, radiative losses, and coronal heating. The effect of Alfven waves on the expanding coronal plasma is also included via the so-called Wentzel-Kramers-Brillouin approximation. The MAST model provides the distribution of the coronal density and temperature that PFSS cannot. The temperature at the lower boundary in this model is 20,000 K (approximately the upper chromosphere) and the transition region is included in the model.  Special techniques are used to broaden the extent of the transition region so that it is resolvable on 3D meshes, but still gives accurate results for the coronal part of the solution. The coronal heating description is empirical and the coronal densities arise entirely from the heating and its interaction with the other terms in the fluid dynamics set of mass, momentum, and energy equations. The simulation results used here employ SDO HMI magnetograms on a given date as input at the inner boundary $r=R_{\odot}$ of the simulation grid. The simulation grid extends from 1 to 30 $R_{\odot}$. The weak points of these MHD simulations are lower grid resolution than in the PFSS model, magnetogram smoothing, and numerical diffusion. The latter can cause magnetic field lines to display unphysical behavior at large heliocentric distances, typically beyond $15~R_{\odot}$. This does not affect our estimates of connectivity lines since we take the intersection point between the Parker Spiral and the MHD integrated line at lower radii, i.e.,  $r\leq 15 R_{\odot}$.

\subsection{Shock front connectivity with the solar surface}
    
  \begin{figure*}
  \centering
  \includegraphics[width=0.8\textwidth]{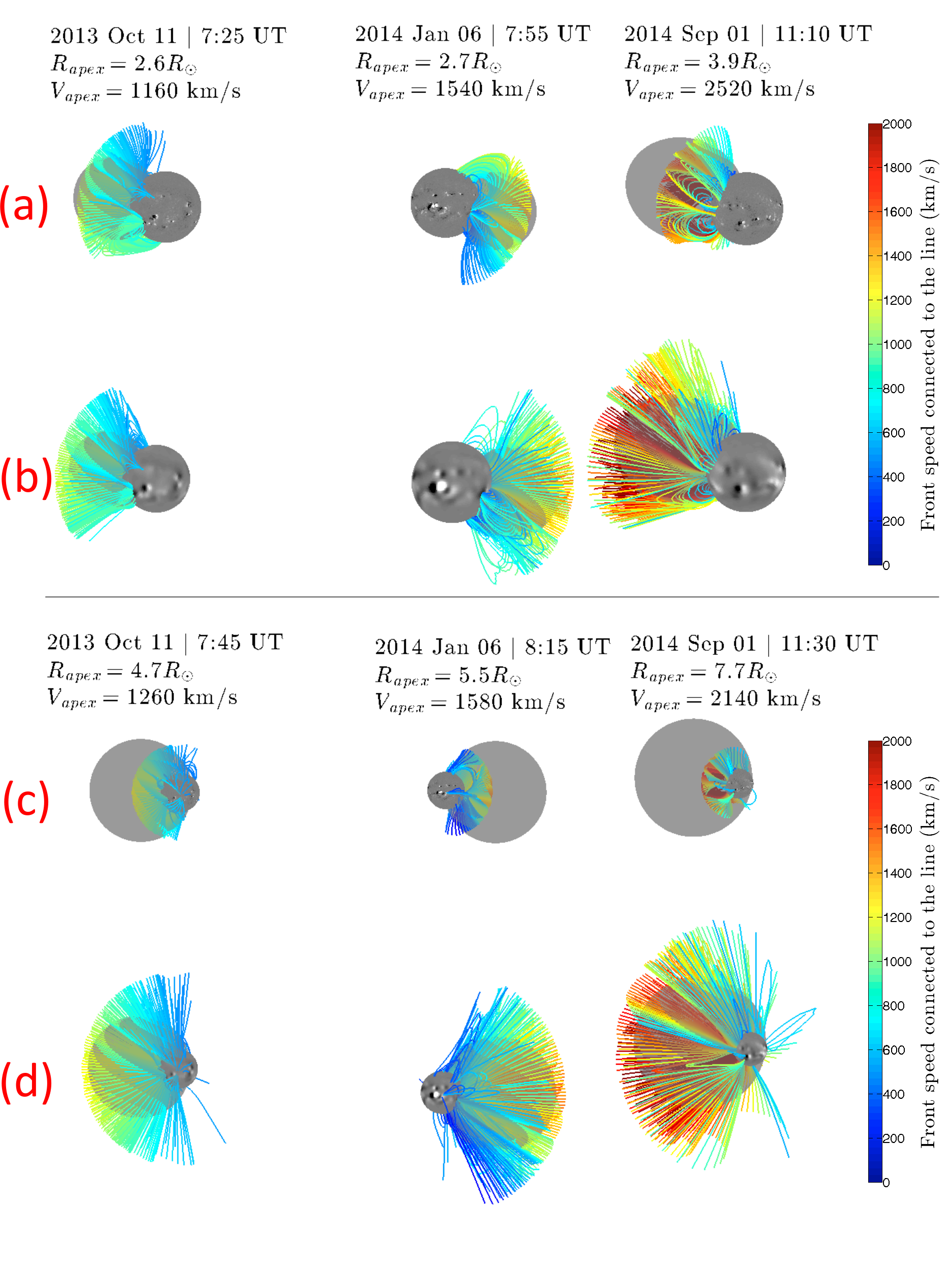}
   \caption{Lines of connectivity between the shock front (smooth gray ellipsoid) and the Sun (SDO images) as viewed from Earth 10 and 30 minutes after the type III radio event for the PFSS (a, c) and MAST (b,d) models for the three events.  The color coding of the lines depicts the shock speed at the point of intersection of the field line with the shock surface. We do not plot the assumed radial field lines above the source surface at 2.5 $R_{\odot}$  for the PFSS model (rows a and c).  }
   \label{fig:lines}%
   \end{figure*}

We combined our reconstruction of the CME pressure front with the magnetic field models to determine the CME velocities when the shock front passes field lines that reach the solar photosphere visible from Earth.  The results of the MAST and PFSS models were provided on a 3D discrete numerical grid in heliocentric coordinates.  We then interpolated the coronal parameters from the model grids to the grid of the surface front that we approximated with an ellipse.  In doing this we followed the procedure described in \citet{Rouillard16}, which uses trilinear interpolations between two grids in spherical coordinates (volume weighting technique).

There are certain approximations with this procedure,  a coronal shock marks the outer boundary of the region influenced by an erupting CME, therefore upstream of the shock location we enter the background solar corona.  We modeled this background corona, the MAST MHD model, or PFSS magnetostatic reconstructions along with some assumptions about the structure of the interplanetary magnetic field.  In this way we were able to estimate the magnetic link between the coronal shock and the spacecraft making in situ measurements. However downstream of the shock, two types of regions are encountered: first, the magnetic piston that produces the coronal shock (the so-called CME core) with its often complex topology involving magnetic flux ropes and, second, the background corona disturbed by the expanding shock wave. The magnetic piston has an important effect on the post-shock coronal plasma. It exerts an important dynamical pressure and compresses the downstream fluid toward the outer parts of the expanding pressure front. Inclusion of the magnetic piston in our model would reconfigure the     magnetic field lines that are currently connecting the nose of the pressure wave to the vicinity of the source region to the flanks of the eruptive structure. There are other processes that become viable with the inclusion of a piston, including magnetic reconnection, which may also reconfigure the magnetic field topology. Our understanding of the detailed topology of CME cores is not sufficiently advanced at this stage to model this region with enough detail for the purpose of  tracking particles. We, therefore, ignored the presence of the piston and model the magnetic connectivity as if the ellipsoidal shock was in an unperturbed background corona both upstream and downstream of the shock.

We considered both the open and closed magnetic field lines and traced these lines from the shock sunward and anti-sunward until reaching a boundary, either at the photosphere or near 1~AU. For magnetic loops, a shock therefore connects at two distinct locations on the photosphere. In Figure~\ref{fig:lines} we present 10 percent of the 5041 open and closed magnetic field lines that we traced through the simulation grids for the  PFSS (rows a and c) and MAST (rows b and d) coronal models for the three events.  The plotted field lines intersect the pressure front (shown in gray) some 10 (a,b) and 30 (c,d) minutes after the onset of type III radio burst.  The color coding of the field lines is defined by the value of the shock speed at the point of intersection of the line with the shock surface with the red lines mostly associated with the fast nose of the CMEs.

Overall, the PFSS and MAST models provide very similar field line connectivities between the shock front and solar surface and, only occasionally,a point on the pressure front connects to solar surface points that are considerably apart. This difference may be due to the differing magnetogram resolutions; the latter is higher for PFSS. In addition, the presence of a source surface in the PFSS model can force a stronger expansion of magnetic flux tubes between $R_{\odot}$ and $2.5 R_{\odot}$ than modeled in MAST. Both effects are important and it is not obvious to know which one is more realistic. We typically find a discrepancy in the field line mapping between the two models of 10 degrees in heliocentric coordinates at the photosphere. We consider this difference acceptable for this work. We used the MAST model, which provides the necessary coronal parameters to derive the properties of the pressure wave in our ensuing analyses.

For both numerical techniques, we typically find that high-speed ($\geq 1000$ km/s) shock fronts are connected to the visible solar disk within 10 minutes after the first appearance of the CME front surface in EUV images. As discussed later,  $\gamma$-ray emission observed by {\it Fermi}-LAT also begins about 10 minutes following the CME onsets for the 2013 Oct and 2014 Sept events. 

\begin{figure*}
   \centering
   \includegraphics[width=0.93\textwidth]{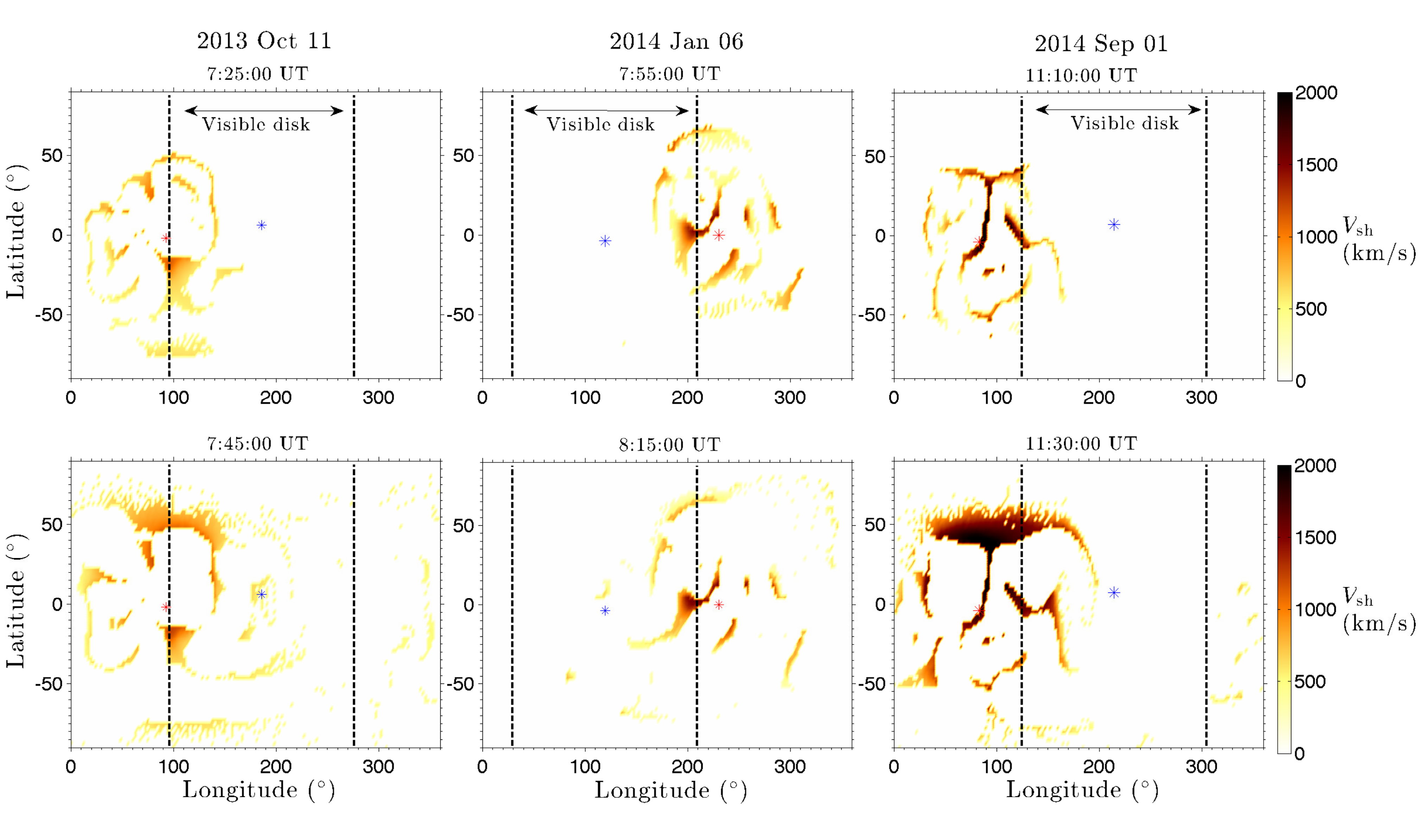}
   \caption{Carrington maps showing the footpoints of magnetic field lines connected to the modeled shock front 10 minutes and 30 minutes after the flare onset derived with the MAST model. Color coding follows the value of the shock front speed $V_{\rm sh}$ magnetically connected to the corresponding solar surface location. The vertical dashed lines mark the approximate location of the solar limbs as viewed from Earth and the blue cross marks the location of the center of the disk. The flare location or equivalently the CME source region is shown as a red cross.}
              \label{fig:Carr_speed}%
    \end{figure*}

A clearer way to present the access of shock-front accelerated particles to the solar atmosphere is to plot the results shown in Figure~\ref{fig:lines} on Carrington maps.  As the MAST and PFSS models give comparable results, and because the MAST model allows us to study more physical aspects of the shock front, we only plot the MAST results in Figure~\ref{fig:Carr_speed}.   The figure presents six Carrington maps showing the location of the photospheric footpoints of magnetic field lines connected to the front/shock surface 10 and 30 min after the CME onsets for the three events. The visible solar disk is delimited by dashed vertical lines and its center is indicated by a blue cross.  For each footpoint we compute the average value of the shock-front speed magnetically connected to that point of the solar surface. The highest front speeds mapping to the visible disk occurred during the 2014 Sept 01 event even though the flare location is the furthest beyond the limb of the three events.  Once again it is clear that shock fronts with speeds $>$1000 km s$^{-1}$ were connected to the visible solar disk within 10 minutes of the CME onsets.

 During the first hour of the CME eruption, the distribution of footpoint locations are associated both with coronal holes where open field lines are rooted and  coronal loops that extend from the solar surface to the shock. As the shock expands the size of the pattern increases and large front speeds connect to an increasing surface area. Even one hour after onset, when the front is connected to field lines intersecting the entire visible Sun, 30\% of the field lines are closed magnetic loops, and not connected to the interplanetary medium.  Most interestingly, within 10-15 minutes of the flare onset, the solar disk visible from Earth is magnetically connected to fast shocks for all three events even though there was a large disparity in speed between the different events that reflect the different kinematic properties already observed in Figure \ref{fig:bulle_apex}.

%\section{Coronal shock evolution and solar $\gamma$-ray emission}
\section{Comparing physical parameters of the shock -- magnetic field-line model with $\gamma$-ray observations} 
\label{sect:evolution}

In addition to the magnetic connectivity it is possible to derive the shock normal angle with respect to the local magnetic field orientation $\theta_{\rm Bn}$ and different Mach numbers (only for the MHD model). For the 2012 May 17 CME, it was shown that high fast-magnetosonic number $M_{fm}>3$ regions can develop on the shock front \citep{Rouillard16}.

As we discussed earlier, the MAST model provides the capability of including physical quantities other than the shock-front speed into the magnetic connection model.  These parameters include the magnetic field strength and orientation, plasma temperature, and density at the front location that enable us to estimate the magnetic field obliquity with respect to the shock-normal $\theta_{\rm Bn}$ and Mach numbers. These parameters are defined by the formulae
        \begin{eqnarray}
        \cos  \theta_{\rm Bn} &=& {\vec{B} \cdot \vec{n} \over |\vec{B}|} \\
        M_A &=& {V_{\rm sh} -\vec{V_w} \cdot \vec{n} \over V_A}\\
         M_{fm} &=& {V_{\rm sh} -\vec{V_w} \cdot \vec{n} \over V_{fm} } ,
        \end{eqnarray}
        where $\vec{n}$ is the local shock-normal vector, $V_{\rm sh}$ is the local shock speed, $V_A$ is the Alvf\'en speed, $V_{fm}$ is the fast-magnetosonic speed, and $\vec{V}_w$ is the background solar wind speed vector. The latter enters in the Mach number estimation because the appropriate frame for our calculations is the local fluid frame, i.e., the background solar wind. The use of the fast magneto-sonic number instead of the Alfv\'enic Mach number is more relevant here because the magnetic field intensity can drop to very low levels in the coronal neutral region where magnetic flux tubes expand greatly and can even annihilate owing to the presence of current sheets. The sound speed remains at about the same value in the quasi-isothermal upper corona. Since $M_{fm}$ retains the contribution from both speeds, we thereby avoid excessively high values of the Mach number in such regions. \\

   \begin{figure*}
   \centering
   \includegraphics[width=0.93\textwidth]{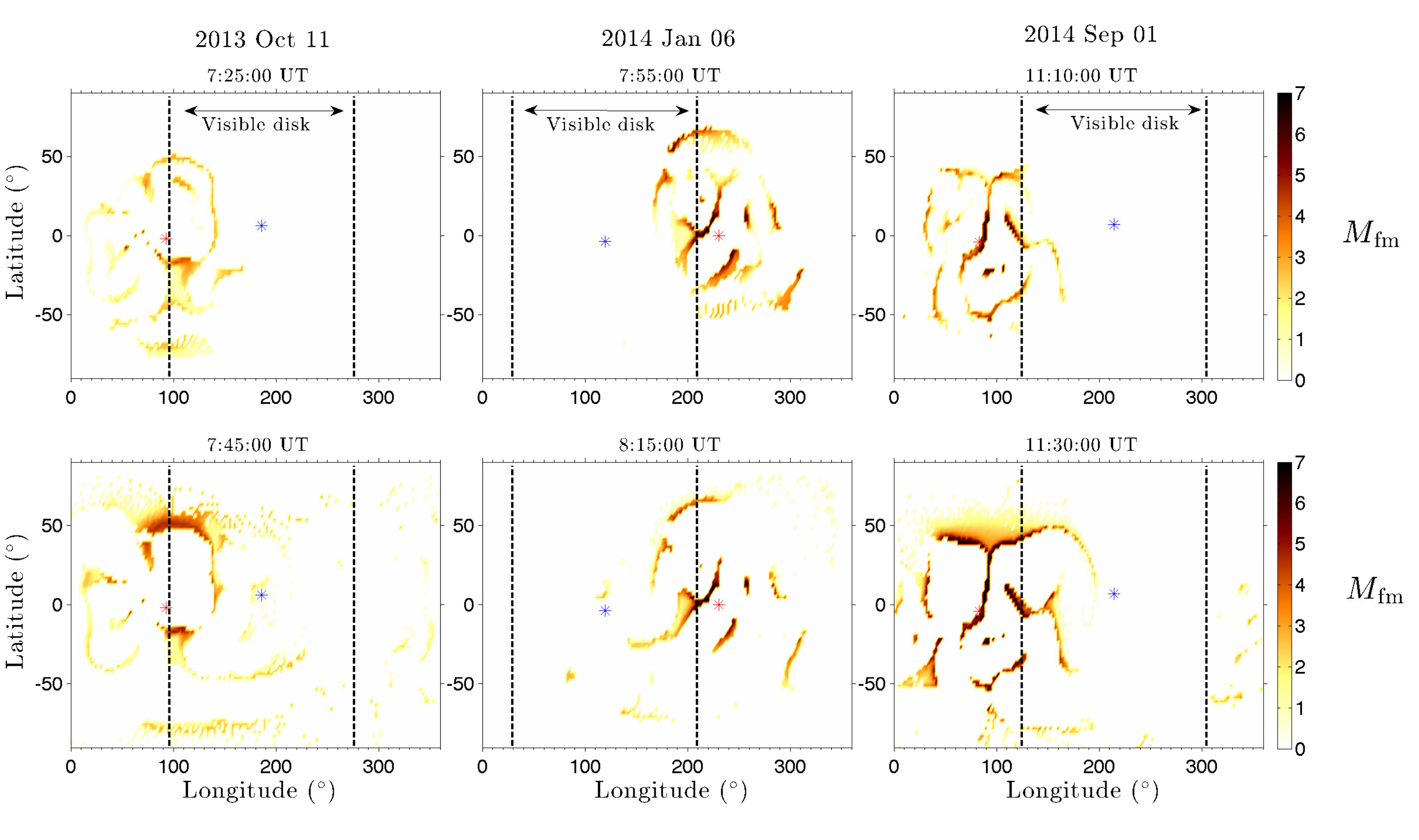}
   \caption{Carrington maps showing in the same format as in Figure~\ref{fig:Carr_speed}, the distribution of  $M_{fm}$. }
              \label{fig:Carr_Mach}%
    \end{figure*}

Shock regions with critical Mach numbers $M_{cr}$, larger than $\simeq 2.7$  are usually considered efficient particle accelerators. Self-consistent, but still idealized simulations, indicate that the particle acceleration efficiency of $M_{fm}>5$ shocks is such that it can transmit more than 10\% of its kinetic energy into energetic ions \citep{2014ApJ...783...91C}. In the precise context of coronal shocks, \citet{2008ApJ...686L.123N} demonstrated that a 2500~km/s quasi-parallel shock propagating through coronal media with typical magnetic fields and sound speeds accelerate seed protons from several keVs to $\sim$GeV energies in 10 minutes.  These authors also showed that both shock speed and Mach number are both important parameters to consider.

We computed $M_{fm}$ over the shock-front surface by dividing our estimated speeds over the 3D ellipsoid with the fast-mode speeds obtained with the plasma parameters modeled by the MAST model. In Figure~\ref{fig:Carr_Mach} we project these local $M_{fm}$ values onto the solar surface by tracing the magnetic field lines connecting the shock to the solar photosphere (as illustrated in Figure~\ref{fig:lines}). The $M_{fm}$ values are plotted onto Carrington maps in the same way as we plotted CME speeds in Figure~\ref{fig:Carr_speed}. We find that the visible disk connects to high $M_{fm}$ values ($\simeq 2.7$) during all three events as early as 10-15 minutes after the CME onset. Unlike the projected shock-front speed, there is no striking difference in the three events with regards to the $M_{fm}$ values projected on the visible disk. Each event has a contribution with $M_{fm}\geq 3$ on the visible disk from early post-flare times. The contribution is most clear for the 2014 Jan 06 event and more marginal for the other two. Assuming that a fast shock with high-Mach number is an efficient particle accelerator, the magnetic connection established early in the event with the visible solar disk may allow these particles to propagate and impact the surface and produce $\gamma$-rays visible from Earth. As previously reported by \citet{Rouillard16} and confirmed for the three events studied here, the highest $M_{fm}$ occurs in the regions where the magnetic field values are very low and the densities are high. This typically corresponds to the location of the coronal neutral current sheet, which is the source region of the heliospheric current sheet and its associated heliospheric plasma sheet. These regions could be efficient particle accelerators for a number of reasons discussed in \citet{Rouillard16}. 

Also, there is an interesting feature when we visually compare Figures \ref{fig:Carr_speed} and \ref{fig:Carr_Mach}. For the regions with roughly equal $V_{\rm sh}$, the highest $M_{fm}$ is reached at the boundaries of the corresponding coronal holes. This feature is logical because magnetic flux tubes expand more rapidly near the coronal hole boundaries than in the central regions. This effect is most striking for the 2014 Sep 01 event at 11:30~UT.

\begin{figure*}
    \centering
     \includegraphics[width=0.95\textwidth]{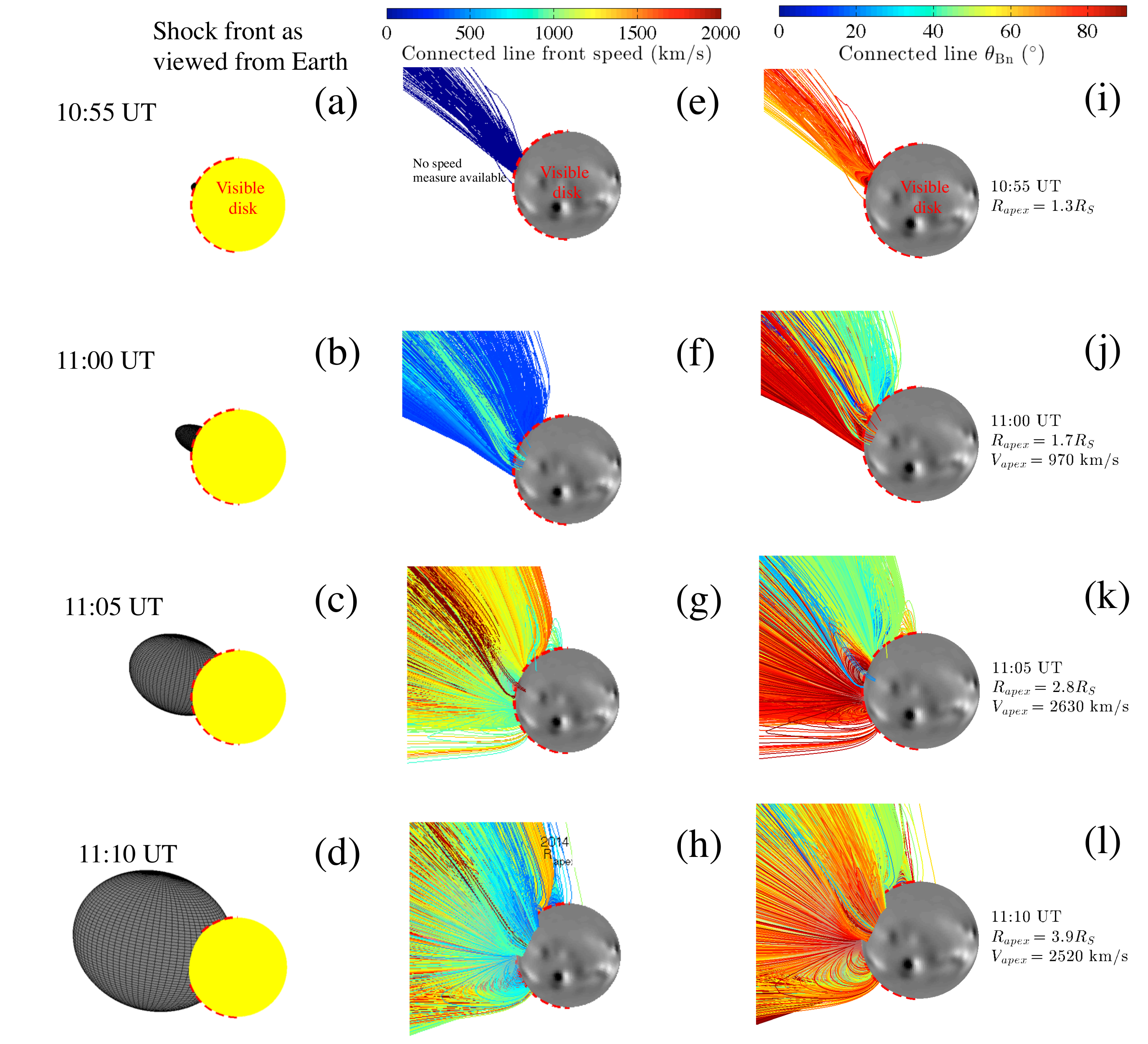}
       \caption{Temporal evolution of the shock front and magnetic connectivity to the solar atmosphere early in the 2014 Sept 01 event derived via the MAST model. Left column (panels (a)-(d)) presents the 3D ellipsoid approximation to the pressure-wave front at 10:55, 11:00, 11:05, and 11:10 UT, respectively,  derived from our CME triangulation studies. The visible solar disk is drawn as a yellow sphere delineated by the red dashed line. The corresponding magnetic field lines color coded by the shock-front speed are plotted in the central column (panels (e)-(h)). The same lines, coded by the value of the magnetic field inclination to the shock-normal $\theta_{\rm Bn}$, are plotted in the right column (panels (i)-(l)). The Sun is represented by a generic magnetogram. The color-coded scales are given at the top of the figure.}
              \label{fig:combi12}%
\end{figure*}

We now look in more detail at the early temporal evolution of the shock connectivity and magnetic geometry in the event on 2014 Sept 01, which had the fastest CME and a source region that was situated furthest behind the limb. In Figure~\ref{fig:combi12} we present the first 15 minutes of shock evolution, in 5 minutes snapshots, as seen from Earth. The left-hand column (panels (a)-(d)) presents the 3D location of the pressure front at 10:55, 11:00, 11:05, and 11:10 UT, respectively. Magnetic field lines, color coded by shock-front speed are plotted in the central column (panels (e)-(h)). The same lines, color coded by the value of the magnetic field inclination to the shock-normal $\theta_{\rm Bn}$ are plotted in the right column (panels (i)-(l)). As can be seen, there was no clear magnetic connection of any fast-moving shock regions to the visible disk before 11:05 UT and no magnetic connection at all at 10:55 UT. This also excludes any magnetic connection of the flare region itself to the visible disk since the shock represents the outermost region disturbed by the CME eruption. The figure also clearly illustrates the quasi-perpendicular nature of the shock (i.e., $\theta_{Bn}>50$ degrees), which is magnetically connected to the visible disk between 11:05 and 11:10 UT. The associated front speeds for these times range from 500 km/s to 1700 km/s with a mean value of $\sim$ 800 km/s and mean Mach $\langle M_{fm} \rangle \simeq 2$. This time range also corresponds to the onset and increase in $\gamma$-ray flux observed by {\it Fermi}-LAT. 

The overall evolution is similar for the two other events: the region of  the early emerging shock front that connected to the visible disk corresponded to the flanks of the pressure wave. Therefore these regions propagated transversely to the  approximately radial coronal field lines or therefore the interaction was predominantly quasi-perpendicular. As shown by \citet{Rouillard16} and recently by \citet{2017ApJ...838...51L},  the mostly parallel region of the CME shock relative to the field lines, at least during the early $<30$ min post-flare evolution, corresponds to its `nose'. The anchor points of the nose-connected field lines for all three events are occulted for an observer at Earth during the early post-flare times. At later times, when the shock front engulfs a significant part of the Sun, $\theta_{\rm Bn}$ evolves gradually to a quasi-parallel configuration.

\begin{figure}
    \centering
     \includegraphics[width=0.46\textwidth]{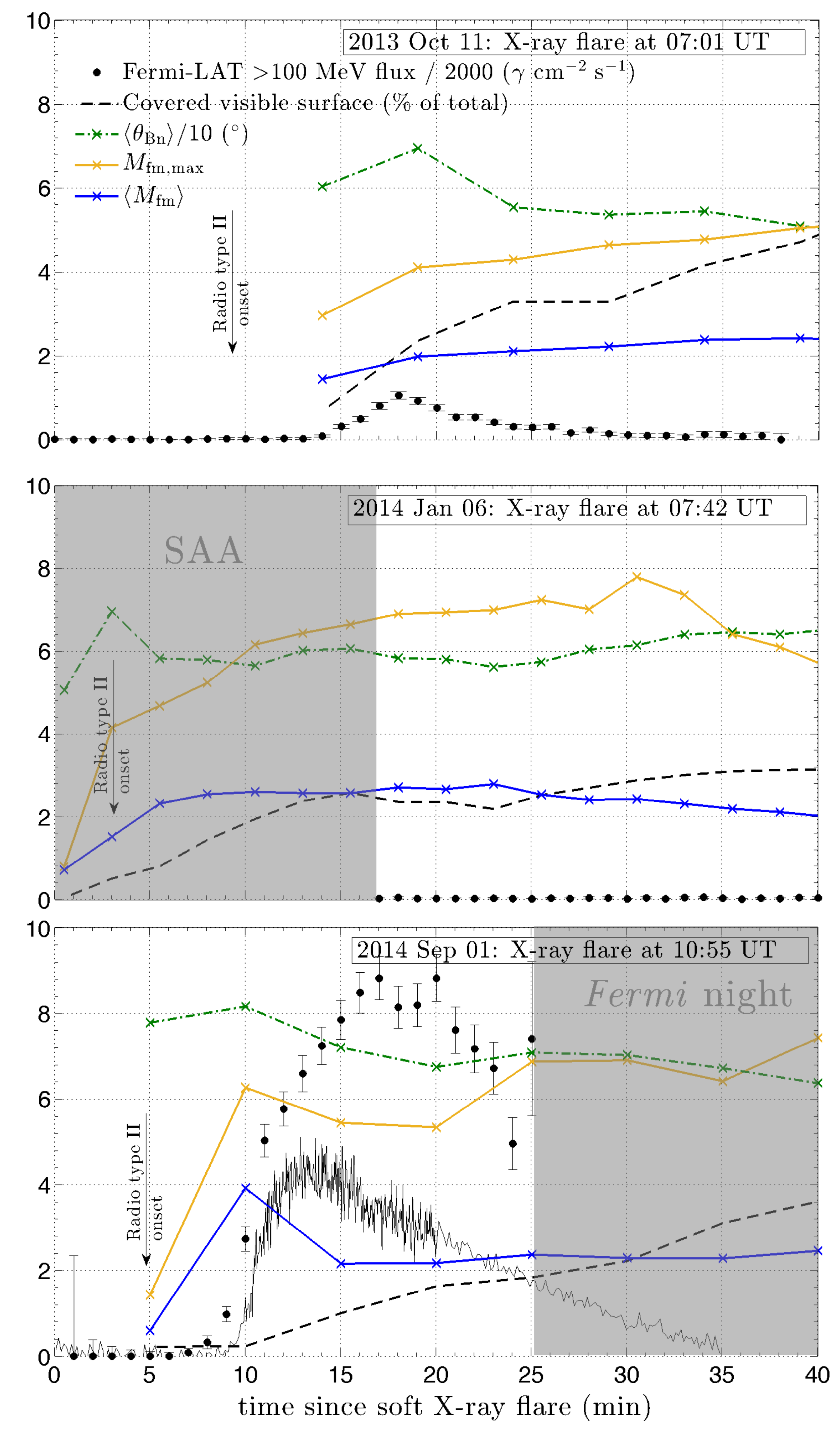}
       \caption{First 40 minutes of the three events: 2013 Oct 11 (top), 2014 Jan 06 (middle), and 2014 Sept 01 (bottom). We show time series of the maximum Mach number ($M_{\rm fm, max}$; orange line), the average Mach number ($\langle M_{fm} \rangle$, blue line), and the average shock geometry $\langle \theta_{\rm Bn} \rangle$ divided by 10 (green dot-dashed line)  derived for all magnetic field lines connected to the visible solar disk. The times at which these quantities where derived are denoted with `x' symbols over the lines. The fraction of the visible solar surface magnetically connected to the shocks is also shown as a dashed black line. Time 0 is defined by the rise in soft X-ray flux. The onset of the metric type II bursts are indicated by vertical arrows and the greater than 100 MeV $\gamma$-ray flux measured by the {\it Fermi}-LAT instrument is shown as filled black circles with corresponding error bars. The $\gamma$-ray fluxes presented in the three panels are all multiplied by factor of 2000 for plotting purposes. Times when the {\it Fermi}-LAT instruments were not observing the Sun are shown as gray shaded areas. In panel (c) the solid black curve represents the arbitrarily scaled 100-300 keV count rates observed by the GBM.}
              \label{fig:last}%
\end{figure}

Figure~\ref{fig:last} compares the maximum Mach number, $M_{\rm fm, max}$, the average Mach numbers, $\langle M_{\rm fm} \rangle$, and the average shock geometry $\langle \theta_{\rm Bn} \rangle$ for the shock front crossing all the magnetic field lines connected with the visible disk for the three events as a function of time using the MAST model. The three panels present the evolution of these quantities for 2013 Oct 11, 2014 Jan 06, and 2014 Sept 01 from top to bottom, respectively. The temporal resolution for determining these parameters is 5 min. The figure also shows the fraction of the visible disk connected to the shock, the flux of 100 MeV $\gamma$ rays measured by {\it Fermi}-LAT, and 100--300 keV hard X-rays for the 2014 Sept 1 event. The periods when {\it Fermi}-LAT did not observe the Sun (South Altlantic Anomaly (SAA) or {\it Fermi} night as described in \citet{2017ApJ...835..219A}) are delimited by gray-shaded areas. The start times of the 2013 Oct 11 and 2014 Sept 1 soft X-ray flares were determined using the SAX instrument on {\it MESSENGER} ($\S$\ref{subsect:flare_oct13} and $\S$\ref{subsect:flare_sept14}).

 The expansion of the coronal pressure wave started at 07:10~UT on 2013 Oct 11, 10 minutes after the X-ray flare. Hence our speed measurements started around 07:15~UT; this explains why the shock-related quantities are delayed compared to the X-ray flare for this event (top panel of the Figure). This delay is less pronounced for the two other events. As shown in this figure early connected parts of the shock to the visible disk are quasi-perpendicular ($\theta_{\rm Bn} > 50^{\circ}$) and super magneto-sonic $\langle M_{fm}>1$ in each case. Some parts  reach supercritical values in each event rapidly after connecting to the visible disk ($M_{rm, max} >3$). The values of these quantities reached early on seem to remain roughly constant at later times.  The mean and maximal values of the Mach number remain stable around 2 and 5, respectively, up to 40 minutes after the start of the soft X-ray flare and $\langle \theta_{\rm Bn}$ decreases slightly from quasi-perpendicular values $>50^{\circ}$ toward quasi-parallel configurations at much later times.  The covered visible disk by shock-connected field lines is very small during the early post-flare times and increases rapidly, up to 5\% values during the bulk of the $\gamma$-ray emission.
 
Even though the time resolution for determining the various shock parameters is only 5 min, we see that the $\gamma$-ray flux increases when the fast shock regions form around the  CME for both the 2013 Oct 11 and Sept 01 events.  This also corresponds to the time when parts of the shock with high $M_{fm}$ values and with quasi-perpendicular geometry become magnetically connected with the solar disk visible from Earth.  This is shown with the increases in the colored curves reaching the visible disk in Figure \ref{fig:combi12}.

The LAT detected $\gamma$-rays near 8:00 UT (18 minutes after the soft X-ray flare onset) on 2014 Jan 06 but the signal is very weak, as discussed in section~\ref{sect:obs}. {\it Fermi} was in the SAA during the first 15 minutes of the event when the CME began. The type II radio burst and our reconstruction suggest that the shock formed during the SAA  transit and that a supercritical section of the shock was connected to the visible disk at that time. These are favorable conditions for the production of $\gamma$-rays.  A weak ground level enhancement was also detected by the South Pole neutron monitor \citep{Thakur14}.  \citet{Thakur14} estimated that the solar release time of relativistic protons occurred at 07:55UT just before LAT detected $\gamma$-rays.
        
 \section{Comparing physical parameters of the shock -- magnetic field-line model with SEP observations} 
 \label{sect:shock_sep}
 \begin{figure}
    \centering
     \includegraphics[width=0.46\textwidth]{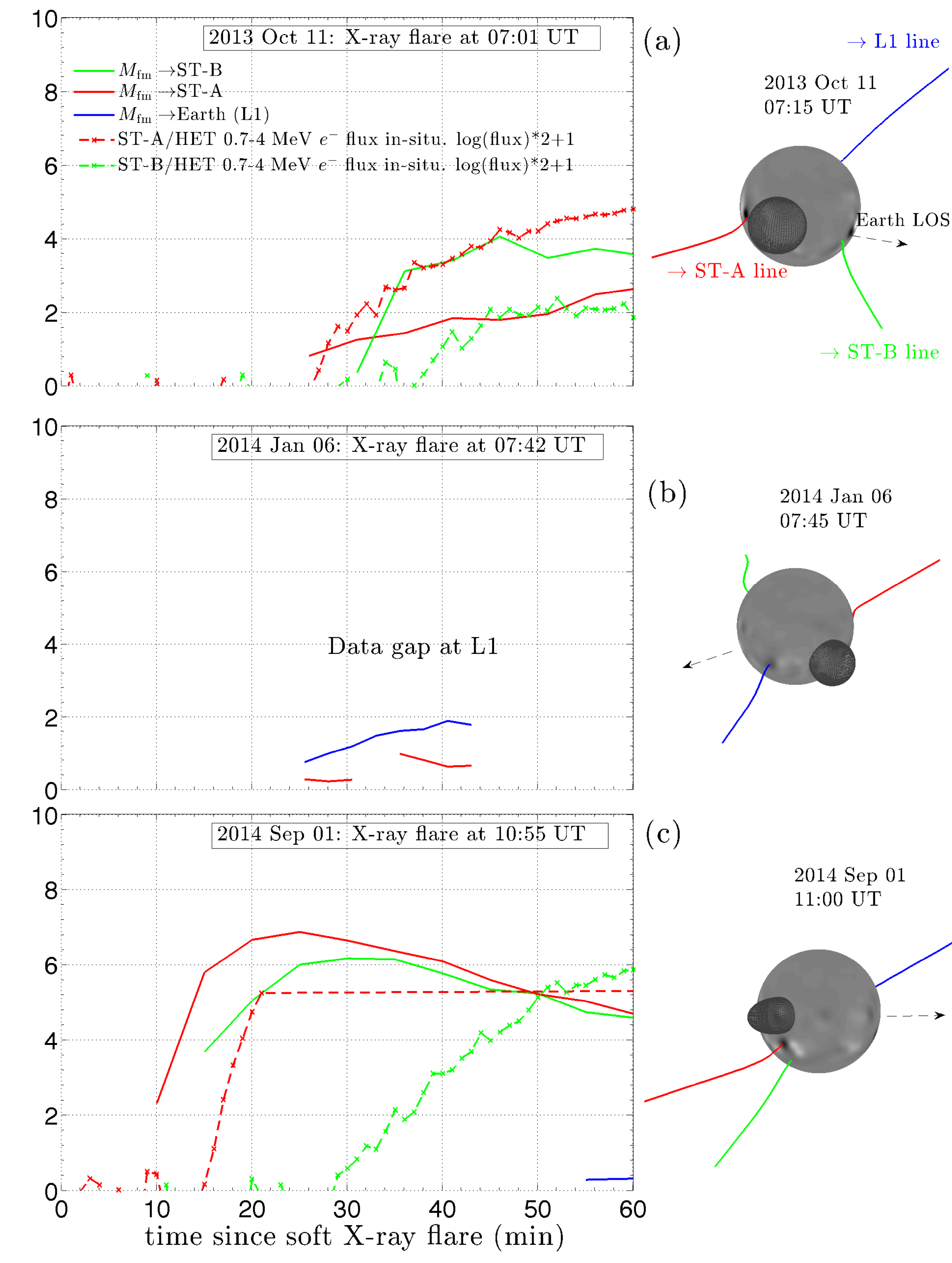}
       \caption{First 60 minutes of the three events, 2013 Oct 11 (panel a), 2014 Jan 06 (panel b) and 2014 Sept 01 (panel c) where we show the time series of Mach numbers at the sites where the shock fronts intersect the field lines crossing the locations of STEREO-A line (solid red line), STEREO-B line (solid green line), and  L1(solid blue line). These times have been corrected for $\sim$ 11 min electron propagation time along the Parker spiral.  The in situ energetic electron fluxes in the $0.7-4$~MeV energy band are plotted with dotted lines: red for {\it STEREO}-A and green for {\it STEREO}-B. Time 0 is defined by the soft-X ray flux onset as observed from {\it MESSENGER}. The right-hand panels show the magnetic connectivity field lines, along with the early post-flare shapes of the coronal shocks, near the times of the radio type II onsets. The dashed black arrows indicate the line of sight as seen from the Earth.}
              \label{fig:shock_sep}%
\end{figure}
   
It is possible that shock-acceleration of particles onto magnetic field lines may be responsible for both the protons interacting on the visible solar disk that produced the $\gamma$-ray emission and the SEPs observed by spacecraft at 1 AU. In the previous section we showed that the onset times of the $\gamma$-rays in two of the events began after the high-Mach shock crossed field lines connected to the visible disk.  In this section we investigate whether the magnetic connectivity of the same CME shock fronts to the {\it STEREO} spacecraft precede the observed onset times of the SEPs discussed in $\S$~\ref{sect:sep_obs} and in appendix~\ref{sec:sep}.
 
On the right side of Figure~\ref{fig:shock_sep} we plot the field lines connecting the Sun to the {\it STEREO} spacecraft and to L1 at about the times of the onset of the type II radio bursts.  Also shown are the ellipsoid models of the developing shock fronts.  The field lines are assumed to follow a Parker spiral between the spacecraft making in situ measurements down to $15~R_{\odot}$. This spiral is defined by the solar wind speed measured in situ and given in Table~\ref{table2}. Below $15~R_{\odot}$  down to the solar surface, we used the MAST MHD model to derive the topology of the field lines.

On the left side of the figure we plot the time series of the shock Mach numbers where the connected field lines intersect the locations of {\it STEREO-A} (solid red line), {\it STEREO-B} (solid green line), and L1 (solid blue line).  These are compared with the time profiles of the flux $0.7-4$~MeV electrons, plotted in (a) for the 2013 Oct 11 and (c) for the 2014 Sep 01 event. We chose the $0.7-4$~MeV energy band because these electrons are relativistic and the most energetic of these electrons propagate almost scatter free from the accelerator to the in situ probe. If the high Mach number shocks accelerated the electrons, we would expect that the intersection times of the shocks with the magnetic field lines would agree with the electron onset times, after taking into account the $\sim$ 11 min electron propagation time along the Parker spiral.

 For the 2013 Oct 11 (a) and 2014 Sep 01 (c) events where direct comparisons were possible, we find that the earliest intersection times of the $M_{\rm fm} >1$ shock regions with field lines connected to {\it STEREO}-A  occur several minutes before the onset times of the SEP electrons at the spacecraft. In addition, the high Mach-number shock connectivity model also predicts the earlier onset times of several minutes for SEP electrons at {\it STEREO}-A than at {\it STEREO}-B.  Unfortunately, there were no energetic electron measurements available for the 2014 Jan 06 event from the Wind or SOHO spacecraft, located at the well-connected L1 point to the early $M_{\rm fm} >1$ shock front.

We therefore conclude that the CME shock-acceleration model accounts for both the onset times of SEP electrons  at 1 AU and the onset times of the $\gamma$-rays produced by protons propagating to the solar atmosphere visible from Earth.

 \section{Summary and discussion}
 \label{sect:discussion}
We have analyzed three CME events on 2013 Oct 11, 2014 Jan 6, and 2014 Sep 1, which (1) erupted behind the solar limb as viewed from Earth, (2) were all associated with the formation of shocks low in the corona as confirmed by the detection of metric type II radio bursts, (3) were associated with detections of $>$100 MeV $\gamma$-rays by the {\it Fermi}-LAT instrument, and (4) were associated with detections of solar energetic particles in the interplanetary medium. Our goal was to test the hypothesis that the shock that formed around the CME could have accelerated the particles responsible for both the $\gamma$-ray emission and the SEPs.

In order to achieve this goal we used coronagraph images of the CMEs from {\it SoHO}-LASCO and at least one of the {\it STEREO} spacecraft (Figure~\ref{fig:orbital_conf}). From these images (Figure~\ref{fig:TRIANG}), we  reconstructed the expansion of the 3D shape of the CME shock and determined the shock front velocities (Figure~\ref{fig:bulle_apex}).  By employing a global simulation of the coronal magnetic field (here MAST), we determined how and when the shock front connected to the solar disk visible from Earth and to the spacecraft measuring SEPs. We then calculated the times when fast and supercritical parts of the shock,  capable of accelerating protons to energies sufficient to produce pion-decay $\gamma$-rays deep in the chromosphere, were magnetically connected to the visible disk.  Our 3D reconstruction analysis showed that the shock fronts of the three events were already visible around the CMEs at the onset times of the metric type II bursts (Figure~\ref{fig:TRIANG}).  In the 2014 Jan 6 and Sep 1 events we confirmed the presence of the shock at type-II onset by showing that the Mach number exceeded super-magnetosonic values at that time; unfortunately we lacked sufficient observations to infer the shock speed at the time of type-II onset for the 2013 Oct 11 event.

In the three studied cases we demonstrated that the shock Mach number underwent a rapid increase to supercritical values within 15 minutes after the type-II onsets. The highest $M_{fm}$ numbers were measured at the shock front nose and near the coronal neutral sheet, just as for the event analyzed in \citet{Rouillard16}. The latter is a region where the magnetic field is low and the density high in MHD simulations. Some of the parts of the shock became magnetically connected to the visible disk before the onset of the $>100$~MeV $\gamma$-ray emission, thus providing a path between the accelerator and visible disk (Figure~\ref{fig:last}). We also showed that these shocks had a quasi-perpendicular geometry during the bulk of the $\gamma$-ray emission. This is true because the flanks of the coronal shock from far-side solar eruptions connecting to the visible disk early in the event propagate across quasi-radial coronal field. Strong quasi-perpendicular shocks can be efficient particle accelerators as long as the upstream coronal magnetic field is sufficiently turbulent and the acceleration time is smaller than in quasi-parallel configuration \citep[e.g.,][]{2005ApJ...624..765G}.

 The $\gamma$-ray and X-ray observations provide information about energetic energetic protons and electrons in the low corona and chromosphere.  The in situ SEP measurements at 1~AU and at Mars provide information about the proton and electron populations escaping from the solar corona.  In the strongest event, on 2014 Sep 01, the 0.7--4 MeV electron and 13--100 MeV proton SEP fluxes rose to peaks just after the flare.  The peak electron-to-proton flux ratio in these energy bands was about 4.5 as measured in these energy bands.  The electron-to-proton flux ratio was about 4.5 as measured in these energy bands. In $\S$\ref{subsect:flare_sept14} we presented evidence that GBM detected sustained electron bremsstrahlung emission in addition to the SGRE observed by LAT.  If the electrons and protons producing the observed bremsstrahlung 
 and $>$100 MeV $\gamma$-ray emission came from the same CME-shock source as the SEPs, we would expect their relative numbers to be comparable. From our fits to the bremsstrahlung spectrum, we estimate that about 3 $\times 10^{33}$ electrons with energies between 0.7--4.0 MeV interacted in the solar atmosphere. Our fits to the pion-decay $\gamma$-ray spectrum provided us with the spectral index and number of protons above a few hundred MeV.  Our comparison of this number with the limit on the number of protons above 40 MeV from neutron-capture line measurements, discussed in $\S$\ref{subsect:flare_sept14}, constrained the proton spectral index at lower energies.  For a proton spectral index of --3.0 from 10 MeV to 200 MeV that steepened to -4.0 above 200 MeV, we estimate that there were about 5 $\times 10^{33}$ protons with energies between 13--100 MeV that interacted at the Sun. The fact that the electron to proton ratios at the Sun and in space agree to within a factor 5 suggests  a common origin for the particles.

The same analysis of the 2013 Oct 11 event shows that the peak pion-decay $\gamma$-ray flux was about 20 times lower than that in the 2014 Sep 01 event.  This is consistent with the 6 to 10 times lower proton flux measured in the Oct 11 event by {\it STEREO} (e.g.,  Appendix~\ref{sec:sep}) relative to Sept 01. The {\it STEREO} 0.7-4.0 MeV electron data from the {\it STEREO A/B} spacecraft for both events (Figure~\ref{fig:sep_1AU}) indicate that the SEP electron flux was between 50 and 100 times higher higher in the 2014 Sept 01 event than it was in the 2013 Oct 11 event. Using this ratio and the measured peak hard X-ray flux in the October event, we estimate that the 100-300 keV peak flux from electron bremsstrahlung would have been too small to have been observed by the GBM.

We have modeled the CME and shock parameters of the 2014 Jan 06 event even though it was barely detected in $\gamma$-rays. The event is also of interest because the coronal shock wave was fast and there is evidence in  neutron-monitor data for production of  protons up to 700~MeV \citep{Thakur14}. One of the possible reasons for the weak $\gamma$-ray emission is that {\it Fermi} was not observing the Sun during the early phases of the coronal shock evolution between 07:45 and 07:55~UT, when the shock reached high Mach numbers  where it could accelerate particles in a quasi-perpendicular geometry onto  field lines reaching the visible disk.  Although the eruption accelerated particles to high energies, it did not produce high levels of SEP fluxes (see Figure~\ref{fig:sep_1AU}). The peak  flux of $20.6-60$~MeV protons was about 10 times less than for the 2013 Oct 11 event and  several orders of magnitude lower than in the 2014 Sep 01 event.  Hence, even if {\it Fermi} had fully observed the Sun during this event, we would have expected the $\gamma$-ray flux to be weak.

We conclude that acceleration of protons by a common coronal shock can account for both the SEPs observed in interplanetary space and sustained $\gamma$-ray emission observed by LAT in the 2013 Oct 11 and 2014 Sept 1 behind-the-limb events. This conclusion is supported by our detailed modeling of the CME shock and particle acceleration onto magnetic field lines both reaching the visible solar disk and interplanetary SEP detectors.  We found that supercritical shocks were capable of accelerating protons of high energy onto the visible disk fast enough to produce the SGRE observed by LAT.  However, we found no clear correlation between the shock Mach number levels and the intensity of the $\gamma$-ray flux measured by LAT, suggesting that the physics is more complicated.  We find that averaging the parameter $\langle n V_{\rm sh}\rangle$ over the shock parts that are connected to the visible disk can grossly reproduce the time-evolution profile of the $>100$~MeV $\gamma$-ray flux if only $M_{fm}>3$ shock parts are selected. In the latter expression $n$ is the fluid density and $V_{\rm sh}$ is the local shock front speed. This apparent empirical relationship needs to be substantiated by physically grounded models. This will potentially allow us to accurately estimate the time-dependent flux and energy of precipitating particles on the solar surface and escaping into the inner heliosphere.

\begin{acknowledgements}
I.P. kindly acknowledges Rui Pinto for providing the PFSS model grid data and for the help with the MSL RAD data extraction. I.P. acknowledges financial support from the HELCATS project under the FP7 EU contract number 606692.  We thank Brian Dennis and Kim Tolbert for providing MESSENGER/SAX data. G. Share acknowledges support from NSF Grant 1156092, NASA Fermi/GI grant GSFC \#71080, and the EU's Horizon 2020 research and innovation program under grant agreement No 637324 (HESPERIA). We acknowledge usage of the tools made available by the plasma physics data center (Centre de Données de la Physique des Plasmas; CDPP; http://cdpp.eu/), the Virtual Solar Observatory (VSO; $http://sdac.virtualsolar.org$), the Multi Experiment Data $\&$ Operation Center (MEDOC; $https://idoc.ias.u-psud.fr/MEDOC$), the French space agency (Centre National des Etudes Spatiales; CNES; https://cnes.fr/fr), and the space weather team in Toulouse (Solar-Terrestrial Observations and Modelling Service; STORMS; https://stormsweb.irap.omp.eu/). This includes the data mining tools AMDA (http://amda.cdpp.eu/) and CLWEB (clweb.cesr.fr/) and the propagation tool (http://propagationtool.cdpp.eu). The \emph{STEREO} \emph{SECCHI} data are produced by a consortium of \emph{RAL} (UK), \emph{NRL} (USA), \emph{LMSAL} (USA), \emph{GSFC} (USA), \emph{MPS} (Germany), \emph{CSL} (Belgium), \emph{IOTA} (France), and \emph{IAS} (France). The \emph{ACE} data were obtained from the \emph{ACE} science center. The WIND data were obtained from the \emph{Space Physics Data Facility}. The MSL RAD data were obtained through the Planetary Data System (PDS).
\end{acknowledgements}

\newpage

\begin{appendix}
\section{In situ SEPs}
\label{sec:sep}

\subsection{In situ measurements near 1~AU}
\begin{figure*}
   \centering
    \includegraphics[width=0.95\textwidth]{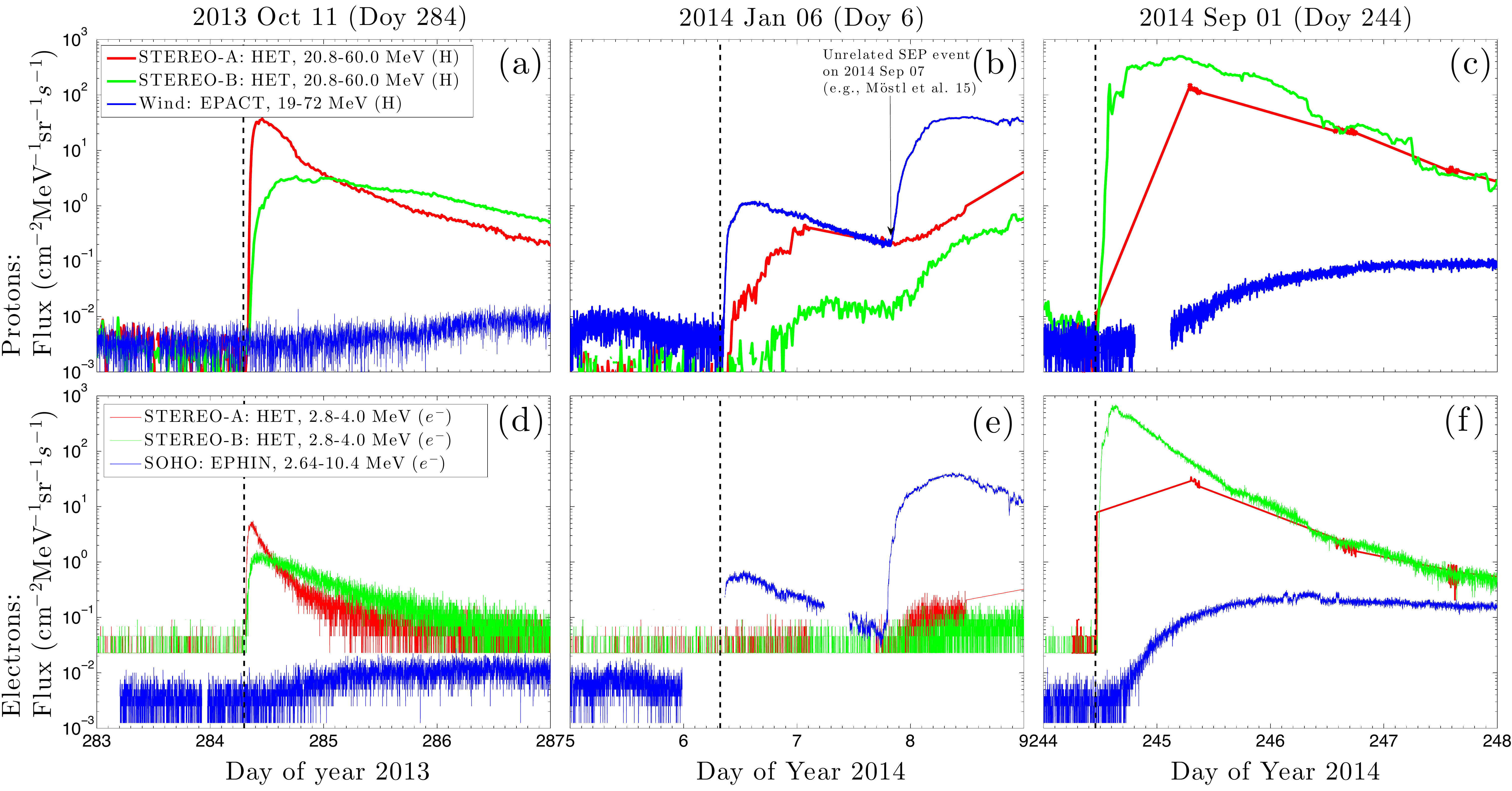}
   \caption{\textit{Top panels}: Time series of proton flux measured by the HET instrument ($20.8-60$ MeV energy band) on {\it STEREO}-A (red lines) and {\it STEREO}-B (green lines) and by the EPACT instrument (19-72 MeV channel) on Wind (blue lines) for the three events: (a) 2013 Oct 11, (b) 2014 Jan 06, and (c) 2014 Sep 01.Vertical dashed lines indicate the type II times as given in Table~\ref{table1}.
   \textit{Bottom panels}: Same as the top panel but for energetic electrons in the $2.8-4.0$~MeV energy band from HET on {\it STEREO}-A (red), {\it STEREO}-B (green), and in $2.64-10.4$~MeV energy band from the EPHIN instrument on SOHO (blue).}
     \label{fig:sep_1AU}%
\end{figure*}

\begin{figure*}
   \centering
    \includegraphics[width=0.85\textwidth]{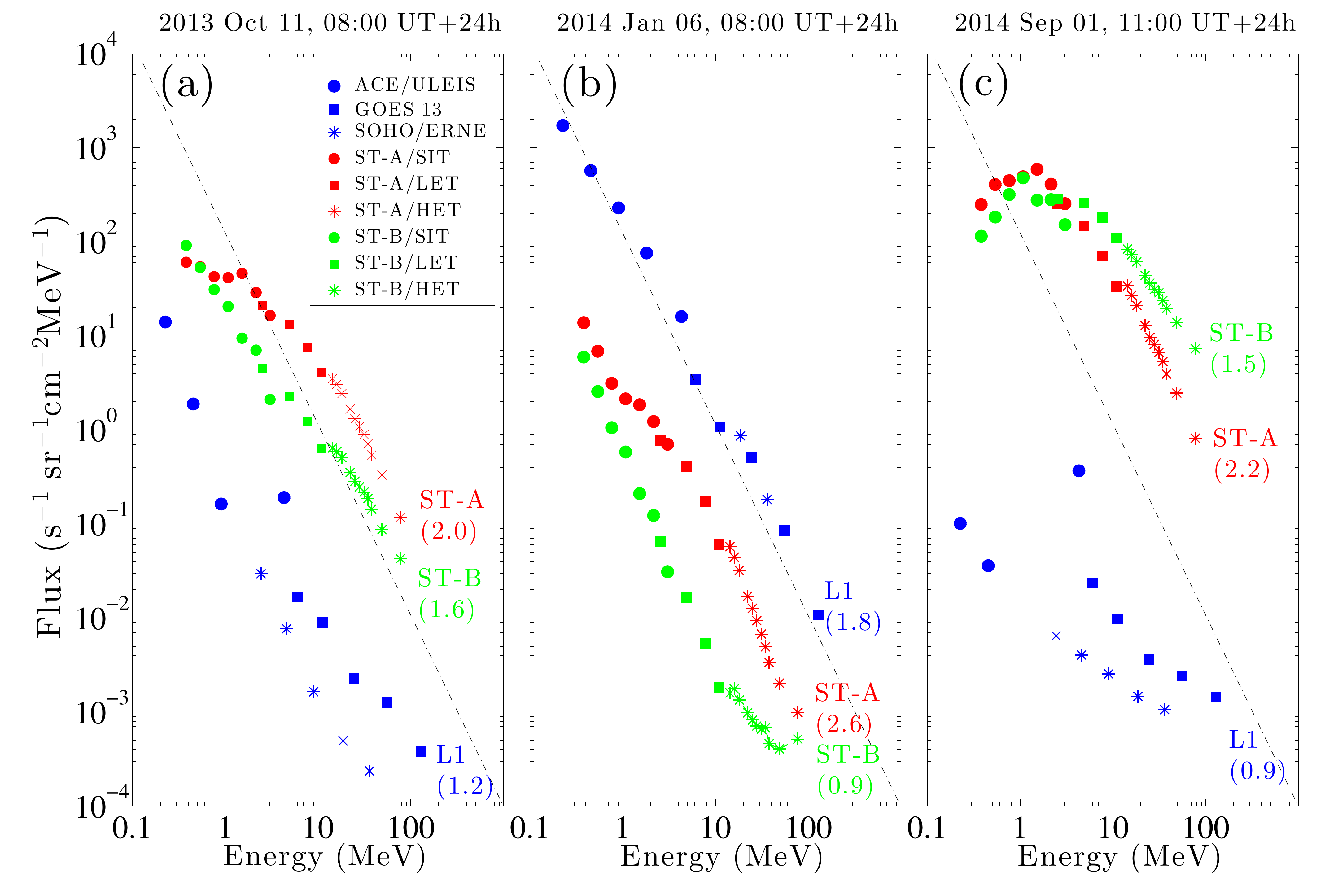}
   \caption{Proton spectra averaged over first 24 hours for the three events at three vantage points  (L1 using blue color symbols, {\it STEREO}-A using red color symbols, and {\it STEREO}-B using green color symbols). Panel (a) represents 2013 Oct 11, panel (b) 2014 Jan 06, and panel (c) 2014 Sep 01. The flux was averaged over the first 24 hours of the SEP event at {\it STEREO}-A and {\it STEREO}-B (SIT, LET, and HET instruments) and L1 (ACE/ULEIS, GOES~13, and SOHO/ERNE instruments). Color coding for different vantage points is the same as in \ref{fig:sep_1AU}. The thin dot-dashed oblique line indicates $\propto E^{-2}$ scaling. }
     \label{fig:sep_spectra}%
\end{figure*}

Figure~\ref{fig:sep_1AU} presents the time evolution of the $20.8-60$~MeV proton flux measurements at {\it STEREO} (HET instruments), the $19-72$~MeV proton flux measurements by the {\it Wind} spacecraft (EPACT instrument; panels a-c), and the time evolution of the $2.8-4.0$~MeV electron flux at {\it STEREO} (HET) and $2.64-10.4$~MeV electron flux by the {\it SoHO} spacecraft (EPHIN instrument; panels d-f). The time series are presented over a four-day interval starting before the eruption and ending several days after. The onsets of energetic protons for the well-connected vantage points occur typically within 30~minutes of the CME onset. This results from either a close connectivity to the flare site or to the shock crossing the probe-connected field line \citep[e.g.,][]{1999SSRv...90..413R, 2012ApJ...752...44R, 2014ApJ...797....8L}. All events last several days and appear to be gradual SEP events.  A small ground-level enhancement was reported by \citet{Thakur14} during the 2014 Jan 06 event (panels (b) and (e)), which was well connected with the Earth, even though it was barely detected in $\gamma$-rays probably because most of the proton interactions occurred on the far side of the Sun.  The largest $20.8-60$~MeV proton flux increase (almost 5 orders of magnitude at {\it STEREO}-B) was observed on 2014 Sep 01 (panels c,f). The peak flux on 2013 Oct 11 (a,d) was one order of magnitude lower (less than 4 orders of magnitude increase at {\it STEREO}-A) while the peak flux for 2014 Jan 06  was only two orders of magnitude above background. This hierarchy seems to follow the observed peak intensities in high-energy $\gamma$-rays observed by {\it Fermi}-LAT.

The temporal evolution of energetic particles flux from different vantage points, as illustrated in Figure~\ref{fig:sep_1AU}, can be summarized as follows. On 2013 Oct 11 {\it STEREO}-A measured a rapid flux increase of relativistic electrons in $2.8-4.0$ MeV band at 07:29~UT and 10 minutes later on {\it STEREO}-B (at 07:40~UT). The peak intensity raised to higher values at {\it STEREO}-A then at {\it STEREO}-B. On L1 only a very gradual increase was measured, starting at $\sim$13:00~UT. On 2014 Jan 06, Earth had good connectivity close to the source region. Hence, energetic proton and electron intensity increased probably within 30 minutes after flare. Yet, electron detectors on {\it SoHO} and {\it Wind} spacecraft do not provide measurements during this early post-flare phase (data gap). Measurements from {\it GOES} \citep{Thakur14, 2017ApJ...835..219A} indicate that $>500$MeV protons were detected starting from 07:57~UT. {\it STEREO}-A and {\it STEREO}-B were probably poorly connected and displayed late increases in energetic protons flux while no significant enhancements are seen in $2.8-4.0$MeV electrons. The SEP event on 2014 Jan 07, visible in (b) and (e), is due to the unrelated flare and CME that occurred close to the center of the visible solar disk \citep{2015NatCo...6E7135M}. Finally, on 2014 Sep 01 both {\it STEREO}-A and {\it STEREO}-B measured a rapid flux increase in $20.8-60$MeV protons and $2.8-4.0$MeV electrons. For electrons the onset was at 11:11~UT and at 11:28~UT on {\it STEREO}-A and {\it STEREO}-B, respectively. Proton onset was measured at 11:48~UT on {\it STEREO}-B while no data point was available from level 2 {\it STEREO}-A data. Indeed, the public STEREO level 2 data excluded some measurements during this period while beacon plots show that, not only the onset occurred earlier on {\it STEREO}-A, but that the peak intensity was higher than on {\it STEREO}-B. This latter is not seen in Figure~\ref{fig:sep_1AU}(c,f) where level 2 data was used. As advised, we used level 2 data here as more relevant for scientific purposes. Concerning the L1 measurements, similarly to 2013 Oct 11 event, a delayed and gradual increase in energetic particles is seen. Following this analysis, Table~\ref{table2} gives the particle onset times from different vantage points and for the three events.

 The broad longitude distribution of the SEPs in space can be tested because at least one of the in situ instruments was magnetically connected to the Sun at a location far from the site of the flare; these are the instruments on the near-Earth spacecraft for the 2013 Oct 11 event, {\it STEREO}-B spacecraft for the 2014 Jan 06 event, and Earth/Mars for the 2014 Sep 01 event. This is shown in the spacecraft connectivity lines in Figure~\ref{fig:bulle_apex}. The broad longitude extent is studied later in this work once the temporal evolution of shock magnetic connectivity is established.

 In Figure~\ref{fig:sep_spectra} we present the proton spectra at the three vantage points (Earth environment, {\it STEREO}-A, and {\it STEREO}-B) and for the three events averaged over 24 hours after SEP onset (start at 08:00 UT, 08:00 UT, and 11:00 UT for 2013 Oct 10 (a), 2014 Jan 06 (b), and 2014 Sep 01 (c), respectively). These spectra were generated using the online OMNIWeb tool that uses the interface developed by Natalia Papitashvili and Joe King\footnote{Tool available at:\\ {http://omniweb.gsfc.nasa.gov/ftpbrowser/flux\_spectr\_m1.html}}. The oblique dot-dashed lines indicate the power-law scaling of energetic particle flux as, $F_{SEP} \propto E^{-s}$ with index $s=2$. The main observation from this Figure is that when the probes were well connected to the flare site they measured a significant increase of the highest SEP flux between 10~MeV and 100~MeV and that the 2014 Sep 01 event produced much higher fluxes of SEPs than the two others. These well-connected vantage points were {\it STEREO}-A and {\it STEREO}-B for 2013 Oct 01 and 2014 Sep 01 events and at L1 for 2014 Jan 06 event. These probes measured SEP flux in 10-100~MeV energy band with spectral indices comprised between 1.5 and 2.2 as indicated in the figure by the values inside brackets, for measurements from near-Earth (blue), {\it STEREO}-A (red), and {\it STEREO}-B (green) spacecraft. This spectral index is close to the generic $s=2$ value. Far connection probes had much lower levels of the average SEP flux and spectral indices can be steeper as well as harder than $s \simeq 2$. At lower energies $<2$ MeV, for 2013 Oct 01 and 2014 Sep 01 events, we observe a plateau in particle fluence, which is possibly due to the trapping of lower energy particles in the vicinity of the shock as a result of efficient wave amplification. This effect was also seen by \citet{2012ApJ...752...44R} in their study of the 2011 March 21 event.

\subsection{Measurements from the Martian surface}

\begin{figure*}
   \centering
  \includegraphics[width=0.98\textwidth]{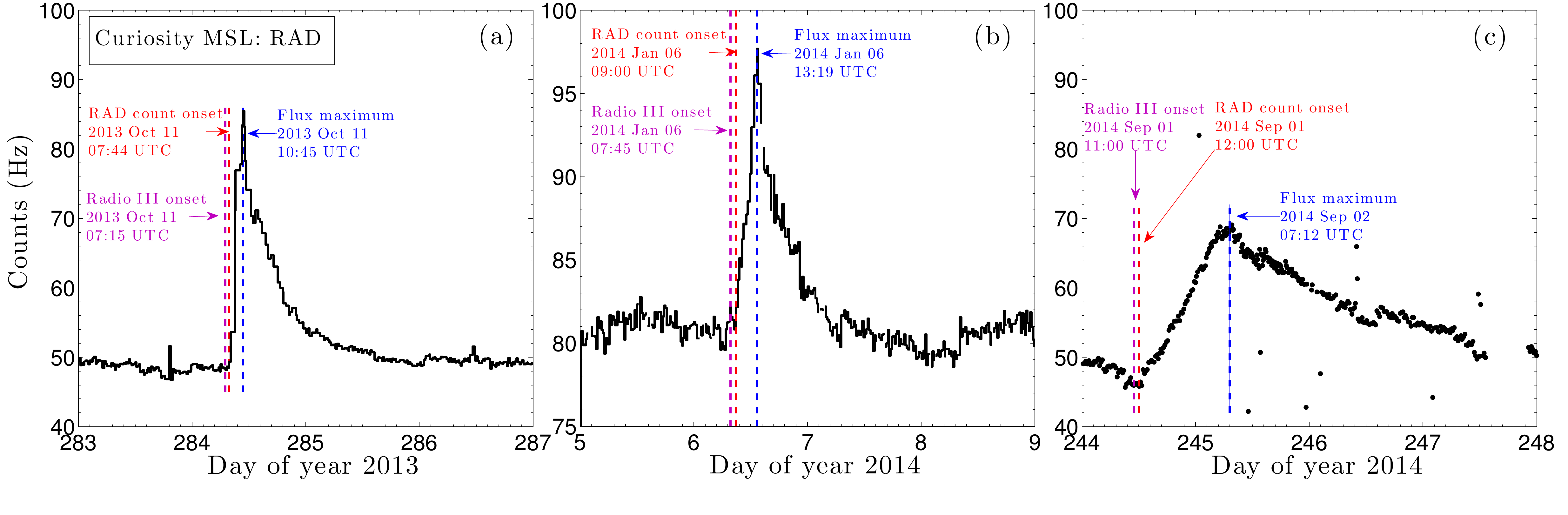}
   \caption{Time series of the counts of energetic particles per second by the RAD instrument on the Mars Science Laboratory's Curiosity Rover for the three events: (a) 2013 Oct 11, (b) 2014 Jan 06 and (c) 2014 Sep 01. Solar radio type III time, particle counts onset time and the time of the peak flux are indicated by the magenta, red and blue vertical dashed lines, respectively.}
    \label{fig:sep_mars}%
\end{figure*}

Measurements from the Radiation Assessment Detector \citep[RAD;][]{2014Sci...343D.386H} on board Martian ground rover Curiosity, i.e., Mars Science Laboratory (MSL) provide additional information on the longitudinal extent of the events, which our model should be able to explain. Here we report on the detection of the energetic particle events by RAD for the three studied events. Figure~\ref{fig:sep_mars} shows the energetic particle counts rate with the same time window as in Figure \ref{fig:sep_1AU}. The times of the solar eruption andthe energetic particles onset are indicated by dashed vertical lines. Mars heliocentric distance was between 1.45~AU and 1.67~AU for the dates of the events. We note that RAD detector is sensitive to $1-1000$MeV protons and $0.2-100$MeV electrons but it requires technical expertise to retrieve the counts from different energy bands and particle species from the public data on the Planetary Data System depository. Here, we plot the ``all included’’ energetic particle count rate, which is equivalent to the dose rate received by the detector. Bearing in mind that we do not distinguish between different particle species we assumed that first arriving particles were either the relativistic electrons or the most energetic protons.  While this procedure provides only limited information compared to the previously discussed 1~AU measurements, we use it here to illustrate the wide longitudinal spread of high particle fluxes measured in the inner heliosphere rapidly after the onset of the solar eruptions.

All three events were clearly detected on the martian surface even though Mars was not always  well connected magnetically to the flare site.    As illustrated in Figure~\ref{fig:bulle_apex}, Mars was well connected to the eruption sites of the 2013 Oct 11 and 2014 Jan 06 events and was poorly connected on 2014 Sep 01 site ($\geq$150 degrees away) (a,b,c). This is reflected in the relatively fast, 3--4 hours, rise times for the two well-connected events and the much longer rise time of the 2014 Sept event.  It is interesting that the particle increases for the three events all started 30 minutes - 1 hour after the accompanying flares. This suggests that some particles had prompt access to the field lines reaching Mars even though the flare site might be over 100 degrees away.

\end{appendix}


\begin{thebibliography}{99}
\bibitem[Ackermann et al.(2012)]{2012ApJS..203....4A} Ackermann, M., Ajello, M., Albert, A., et al.\ 2012, \apjs, 203, 4
\bibitem[Ackermann et al.(2014)]{acke14} Ackermann, M., Ajello, M., Albert, A., et al.\ 2014, \apj, 787, 15
\bibitem[Ackermann et al.(2017)]{2017ApJ...835..219A} Ackermann, M., Allafort, A., Baldini, L., et al.\ 2017, \apj, 835, 219
\bibitem[Afanasiev et al.(2015)]{2015A&A...584A..81A} Afanasiev, A., Battarbee, M., \& Vainio, R.\ 2015, \aap, 584, A81 
\bibitem[Ajello et al.(2014)]{ajel14} Ajello, M., Albert, A., Allafort, A., et al.\ 2014, \apj, 789, 20 
\bibitem[Akimov et al.(1996)]{1996SoPh..166..107A} Akimov, V.~V., Ambro{\v z}, P., Belov, A.~V., et al.\ 1996, \solphys, 166, 107
\bibitem[Aschwanden(2012)]{2012SSRv..171....3A} Aschwanden, M.~J.\ 2012, \ssr, 171, 3 
 \bibitem[Atwood et al.(2009)]{atwo09} Atwood, W.~B., Abdo, A.~A., Ackermann, M., et al.\ 2009, \apj, 697, 1071
 \bibitem[Benz(2008)]{2008LRSP....5....1B} Benz, A.~O.\ 2008, Living Reviews in Solar Physics, 5, 1
 \bibitem[Berezhko \& Taneev(2003)]{2003AstL...29..530B} Berezhko, E.~G., \& Taneev, S.~N.\ 2003, Astronomy Letters, 29, 530 
\bibitem[Blandford \& Eichler(1987)]{1987PhR...154....1B} Blandford, R., \& Eichler, D.\ 1987, \physrep, 154, 1 
\bibitem[Caprioli \& Spitkovsky(2014)]{2014ApJ...783...91C} Caprioli, D., \& Spitkovsky, A.\ 2014, \apj, 783, 91
\bibitem[Chen et al.(2015)]{2015Sci...350.1238C} Chen, B., Bastian, T.~S., Shen, C., et al.\ 2015, Science, 350, 1238 
\bibitem[Chertok(1996)]{1996R&QE...39..940C} Chertok, I.~M.\ 1996, Radiophysics and Quantum Electronics, 39, 940 
\bibitem[Chupp \& Ryan(2009)]{2009RAA.....9...11C} Chupp, E.~L., \& Ryan, J.~M.\ 2009, Research in Astronomy and Astrophysics, 9, 11 
\bibitem[Cliver et al.(1993)]{Cliver93} Cliver, E.~W., Kahler, S.~W., \& Vestrand, W.~T.\ 1993, International Cosmic Ray Conference, 3, 91
\bibitem[Cliver(2016)]{2016ApJ...832..128C} Cliver, E.~W.\ 2016, \apj, 832, 128 
\bibitem[Dresing et al.(2012)]{2012SoPh..281..281D} Dresing, N., G{\'o}mez-Herrero, R., Klassen, A., et al.\ 2012, \solphys, 281, 281 
\bibitem[Ellison \& Ramaty(1985)]{1985ApJ...298..400E} Ellison, D.~C., \& Ramaty, R.\ 1985, \apj, 298, 400 
\bibitem[Forrest et al.(1981)]{1981ICRC...10....5F} Forrest, D.~J., Chupp, E.~L., Ryan, M.~M., et al.\ 1981, International Cosmic Ray Conference, 10, 5 
\bibitem[Giacalone(2005)]{2005ApJ...624..765G} Giacalone, J.\ 2005, \apj, 624, 765
\bibitem[Gopalswamy(2003)]{Gopalswamy03} Gopalswamy, N.\ 2003, \grl, 30, 8013 
\bibitem[Hassler et al.(2014)]{2014Sci...343D.386H} Hassler, D.~M., Zeitlin, C., Wimmer-Schweingruber, R.~F., et al.\ 2014, Science, 343, 1244797 
\bibitem[Hudson(2011)]{2011SSRv..158....5H} Hudson, H.~S.\ 2011, \ssr, 158, 5 
\bibitem[Hurford et al.(2003)]{2003ApJ...595L..77H} Hurford, G.~J., Schwartz, R.~A., Krucker, S., et al.\ 2003, \apjl, 595, L77 
\bibitem[Kaiser et a.l(2008)]{2008SSRv..136....5K} Kaiser, M.~L., Kucera, T.~A., Davila, J.~M., et al.\ 2008, \ssr, 136, 5 
\bibitem[Kanbach et al.(1993)]{1993A&AS...97..349K} Kanbach, G., Bertsch, D.~L., Fichtel, C.~E., et al.\ 1993, \aaps, 97, 349 
\bibitem[Klein et al.(2014)]{2014A&A...572A...4K} Klein, K.-L., Masson, S., Bouratzis, C., et al.\ 2014, \aap, 572, A4 
\bibitem[Klein \& Trottet(2001)]{2001SSRv...95..215K} Klein, K.-L., \& Trottet, G.\ 2001, \ssr, 95, 215 
\bibitem[Kozarev et al.(2015)]{2015ApJ...799..167K} Kozarev, K.~A., Raymond, J.~C., Lobzin, V.~V., \& Hammer, M.\ 2015, \apj, 799, 167 
\bibitem[Lario et al.(2017)]{2017ApJ...838...51L} Lario, D., Kwon, R.-Y., Richardson, I.~G., et al.\ 2017, \apj, 838, 51
\bibitem[Lario et al.(2014)]{2014ApJ...797....8L} Lario, D., Raouafi, N.~E., Kwon, R.-Y., et al.\ 2014, \apj, 797, 8  
\bibitem[Lario et al.(1998)]{1998ApJ...509..415L} Lario, D., Sanahuja, B., \& Heras, A.~M.\ 1998, \apj, 509, 415 
\bibitem[Lee(2005)]{2005ApJS..158...38L} Lee, M.~A.\ 2005, \apjs, 158, 38 
\bibitem[Lemen et al.(2012)]{2012SoPh..275...17L} Lemen, J.~R., Title, A.~M., Akin, D.~J., et al.\ 2012, \solphys , 275, 17 
\bibitem[Lin et al.(2002)]{2002SoPh..210....3L} Lin, R.~P., Dennis, B.~R., Hurford, G.~J., et al.\ 2002, \solphys, 210, 3 
\bibitem[Lionello et al.(2009)]{2009ApJ...690..902L} Lionello, R., Linker, J.~A., \& Miki{\'c}, Z.\ 2009, \apj, 690, 902
\bibitem[Litvinenko(1996)]{1996ApJ...462..997L} Litvinenko, Y.~E.\ 1996, \apj, 462, 997 
 \bibitem[Marcowith et al.(2016)]{2016RPPh...79d6901M} Marcowith, A., Bret, A., Bykov, A., et al.\ 2016, Reports on Progress in Physics, 79, 046901 
 \bibitem[Mewaldt et al.(2008)]{Mewaldt08} Mewaldt, R.~A., Cohen, C.~M.~S., Giacalone, J., et al.\ 2008, American Institute of Physics Conference Series, 1039, 111 
\bibitem[Morlino \& Caprioli(2012)]{2012A&A...538A..81M} Morlino, G., \& Caprioli, D.\ 2012, \aap, 538, A81 
\bibitem[M{\"o}stl et al.(2015)]{2015NatCo...6E7135M} M{\"o}stl, C., Rollett, T., Frahm, R.~A., et al.\ 2015, Nature Communications, 6, 7135 
\bibitem[Murphy et al.(1987)]{1987ApJS...63..721M} Murphy, R.~J., Dermer, C.~D., \& Ramaty, R.\ 1987, \apjs, 63, 721 
\bibitem[Ng \& Reames(2008)]{2008ApJ...686L.123N} Ng, C.~K., \& Reames, D.~V.\ 2008, \apjl, 686, L123 
\bibitem[Patsourakos \& Vourlidas(2009)]{2009ApJ...700L.182P} Patsourakos, S., \& Vourlidas, A.\ 2009, \apjl, 700, L182 
\bibitem[Pesce-Rollins et al.(2015a)]{2015ApJ...805L..15P} Pesce-Rollins, M., Omodei, N., Petrosian, V., et al.\ 2015, \apjl, 805, L15 
\bibitem[Pesce-Rollins et al.(2015b)]{2015arXiv150704303P} Pesce-Rollins, M., Omodei, N., Petrosian, V., et al.\ 2015, arXiv:1507.04303 
\bibitem[Petrosian(2012)]{2012SSRv..173..535P} Petrosian, V.\ 2012, \ssr, 173, 535 
\bibitem[Ramaty et al.(1987)]{1987ApJ...316L..41R} Ramaty, R., Murphy, R.~J., \& Dermer, C.~D.\ 1987, \apjl, 316, L41 
\bibitem[Reames(1999)]{1999SSRv...90..413R} Reames, D.~V.\ 1999, \ssr, 90, 413 
\bibitem[Reames et al.(1996)]{1996ApJ...466..473R} Reames, D.~V., Barbier, L.~M., \& Ng, C.~K.\ 1996, \apj, 466, 473
\bibitem[Rouillard et al.(2012)]{2012ApJ...752...44R} Rouillard, A.~P., Sheeley, N.~R., Tylka, A., et al.\ 2012, \apj, 752, 44 
\bibitem[Rouillard et al.(2016)]{Rouillard16} Rouillard, A.~P., Plotnikov, I., Pinto, R.~F., et al.\ 2016, \apj, 833, 45 
\bibitem[Ryan(2000)]{Ryan00} Ryan, J.~M.\ 2000, \ssr, 93, 581 
\bibitem[Scherrer et al.(2012)]{2012SoPh..275..207S} Scherrer, P.~H., Schou, J., Bush, R.~I., et al.\ 2012, \solphys, 275, 207 
\bibitem[Schlemm et al.(2007)]{schl07} Schlemm, C.~E., Starr, R.~D., Ho, G.~C., et al.\ 2007, \ssr, 131, 393 
\bibitem[Schrijver \& De Rosa(2003)]{2003SoPh..212..165S} Schrijver, C.~J., \& De Rosa, M.~L.\ 2003, \solphys, 212, 165 
\bibitem[Thakur et al.(2014)]{Thakur14} Thakur, N., Gopalswamy, N., Xie, H., et al.\ 2014, \apjl, 790, L13 
\bibitem[Vilmer et al.(2011)]{2011SSRv..159..167V} Vilmer, N., MacKinnon, A.~L., \& Hurford, G.~J.\ 2011, \ssr, 159, 167 
\bibitem[Vourlidas \& Ontiveros(2009)]{2009AIPC.1183..139V} Vourlidas, A., \& Ontiveros, V.\ 2009, American Institute of Physics Conference Series, 1183, 139 
\bibitem[Wang \& Sheeley(1992)]{1992ApJ...392..310W} Wang, Y.-M., \& Sheeley, N.~R., Jr.\ 1992, \apj, 392, 310
\bibitem[Zank et al.(2000)]{2000JGR...10525079Z} Zank, G.~P., Rice, W.~K.~M., \& Wu, C.~C.\ 2000, \jgr, 105, 25079 
\end{thebibliography}
\end{document}